\pdfoutput=1
\documentclass[sigconf,10pt,nonacm]{acmart}

\renewcommand\footnotetextcopyrightpermission[1]{} 

\usepackage{listings}
\usepackage{algorithm}
\usepackage{algpseudocode}
\usepackage{amsmath}
\usepackage{amsfonts}
\usepackage{ulem}
\usepackage{subcaption}
\usepackage{caption}   
\usepackage{adjustbox} 
\usepackage{soul}

\DeclareMathOperator*{\abs}{abs}

\newcommand{\showsolutions}{\long\def\soln##1\solnend{##1}}

\showsolutions

\AtBeginDocument{%
  }

\begin{document}

\title{mCardiacDx: Radar-Driven Contactless Monitoring and Diagnosis of Arrhythmia}


 \author{Arjun Kumar}
 \orcid{1234-5678-9012}
 \affiliation{
   \institution{KAIST}
   \country{Korea}
}
 \email{arjun@kaist.ac.kr}
 \author{Noppanat Wadlom}
 \affiliation{%
  \institution{KAIST}
  \country{Korea}
} 
 \email{noppanat_w@kaist.ac.kr}
\author{Jaeheon Kwak}
 \affiliation{%
\institution{KAIST}
  \country{Korea}
} 
 \email{0jaehunny0@gmail.com}
\author{Si-Hyuck Kang}
\affiliation{%
\institution{Seoul National University Bundang Hospital}
  \country{Korea}
} 
\email{eandp303@gmail.com}
\author{Insik Shin}
\affiliation{%
  \institution{KAIST}
  \country{Korea}
} 
\email{insik.shin@kaist.ac.kr}
%
\begin{abstract}
Arrhythmia is a common cardiac condition that can precipitate severe complications without timely intervention. While continuous monitoring is essential for timely diagnosis, conventional approaches such as electrocardiogram and wearable devices are constrained by their reliance on specialized medical expertise and patient discomfort from their contact nature. Existing contactless monitoring, primarily designed for healthy subjects, face significant challenges when analyzing reflected signals from arrhythmia patients due to disrupted spatial stability and temporal consistency. 

In this paper, we introduce \textbf{mCardiacDx}, a radar-driven contactless system that accurately analyzes reflected signals and reconstructs heart pulse waveforms for arrhythmia monitoring and diagnosis. The key contributions of our work include a novel \textit{precise target localization (PTL)} technique that locates reflected signals despite spatial disruptions, and an \textit{encoder-decoder model} that transforms these signals into HPWs, addressing temporal inconsistencies. Our evaluation on a large dataset of healthy subjects and arrhythmia patients shows that both \textbf{mCardiacDx} and \textit{PTL} outperform state-of-the-art approach in arrhythmia monitoring and diagnosis, also demonstrating improved performance in healthy subjects.
\end{abstract}



\received{20 February 2007}
\received[revised]{12 March 2009}
\received[accepted]{5 June 2009}

\maketitle
\section{Introduction}
 \label{sec:intro}
Cardiovascular diseases (CVDs) are the leading cause of death worldwide, claiming an estimated 20.5 million lives each year \cite{who2024,worldheart2024,whf2023}. Arrhythmia, characterized by abnormal heart activity \cite{xiao2011cardiac}, is one of the most common types of CVDs \cite{xiao2011cardiac}. If left untreated, it can lead to severe complications, such as blood clots, stroke, or even sudden cardiac death \cite{jnj2024}. Therefore, regular monitoring and early diagnosis of arrhythmia are critical to mitigate the risk of adverse events and provide a reference for effective clinical care \cite{calkins2009treatment}.

Given the importance, arrhythmia diagnosis technologies are rapidly growing, particularly with the integration of artificial intelligence into diagnostic tools. While conventional diagnostic tools like electrocardiogram (ECG) in medical institutions are highly effective, their restricted accessibility and the need for specialized medical expertise limit their utility for daily, continuous monitoring. To address this need, ambulatory ECG device such as Holter monitor is commonly used \cite{hannun2019cardiologist,turakhia2013diagnostic}. However, it requires electrode attachment for extended periods, which largely interfere with daily activities. More convenient wearables like smartwatches and wristbands \cite{guo2019mobile,lopez2017oral,perez2019large,zhang2021towards} are gaining attention for their potential in arrhythmia diagnosis. However, a recent study highlights user discomfort during their prolonged use \cite{jeong2017smartwatch}.

\begin{figure}[ht]
  \centering
  \begin{tabular}{cc}
    \includegraphics[width=0.46\linewidth, height=0.40\linewidth]{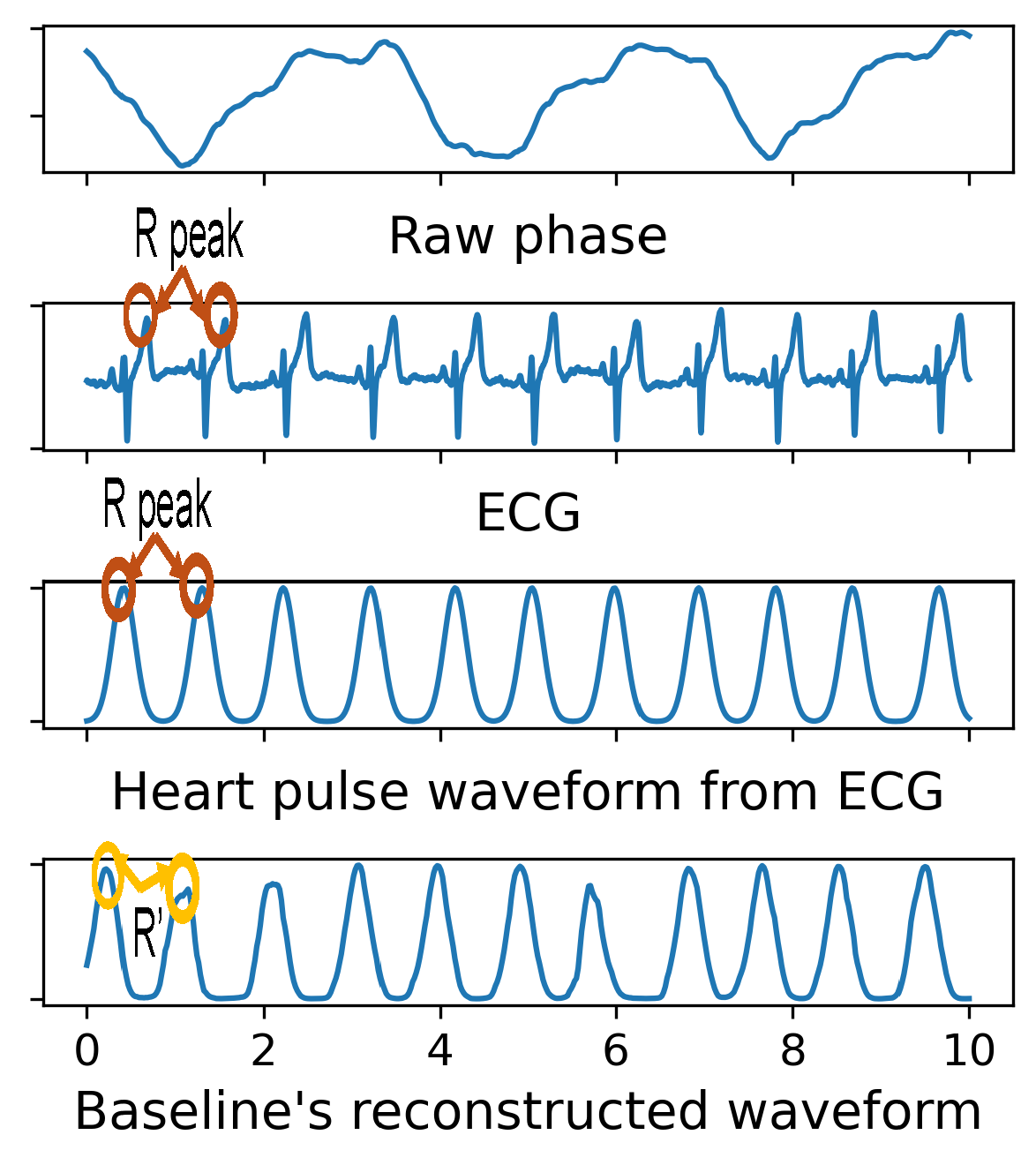} & 
    \includegraphics[width=0.46\linewidth, height=0.40\linewidth]{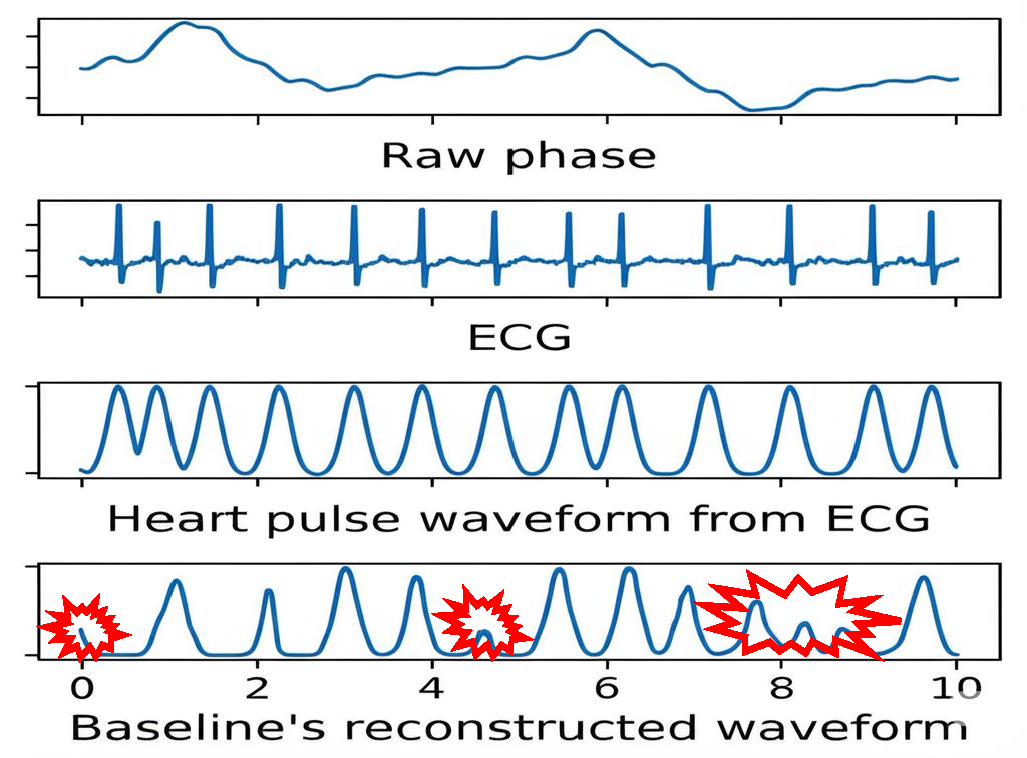} \\ 
    (a) Healthy Subject & (b) Arrhythmia Patient \\
  \end{tabular}
  \caption{Performance of the Baseline.}
  \label{fig:baseline_wave}
    \Description{}
\end{figure}

In response to these limitations, contactless sensing has attracted growing interest from both industry and academia in which radio-frequency (RF) sensing leveraging commercial-grade radars has emerged as a promising solution \cite{}. They focus on monitoring heart rate (HR) \cite{hao2025fmcw,chen2021movi,wang2021mmhrv,zhao2016emotion} and have recently progressed in reconstructing cardiac waveforms such as Seismocardiogram (SCG) \cite{ha2020contactless}, ECG \cite{chen2022contactless,xu2021cardiacwave,wang2023ecg}, and Photoplethysmogram (PPG) \cite{zhang2022can,khan2022contactless}. However, all of these approaches analyze reflection samples from the chest wall vibration \cite{he2009learning,ning2021rf} of healthy subjects. Therefore, these approaches are likely to face significant challenges in arrhythmia patients, because arrhythmia may disrupt the chest wall vibration patterns, which directly impacts the reflection samples in both the spatial and temporal domains.

To better understand the damaging effect of arrhythmia on reconstructing heart pulse waveforms (HPWs) representing heartbeats (R'), similar to HPWs generated from an ECG (R-peaks), we compare two cases: (a) a healthy subject and (b) an arrhythmia patient, as shown in Figure~\ref{fig:baseline_wave}. Even with the state-of-the-art  approach (Baseline) \cite{ha2020contactless}, the arrhythmia-disrupted reflection samples in case (b) fail to yield correct HPWs compared to case (a)  (as highlighted in red). This failure leads to significantly higher median absolute percentage errors (MedAPE) \cite{mdape,sktimeMAPE} of 9.10\% and 8.42\% for HR and RR intervals, respectively, in case (b), compared to much lower errors of 2.66\% and 2.73\% in case (a), all computed relative to the ECG.

We investigated the underlying reasons for such deteriorated results under arrhythmia and observe two fundamental differences. First, high-magnitude reflection samples in healthy case (a) are concentrated into a single, stable range bin corresponding to a strong chest wall vibration at a single spatially stable region. In contrast, arrhythmia in case (b) disrupts this spatial stability, causing reflection samples of varying magnitudes to be dispersed across multiple unstable range bins. This corresponds to multiple spatially unstable  regions on the chest wall, each with varying vibration strength (detailed in Section \ref{c1} (C1)). This spatial dispersion challenges the Baseline to precisely locate the relevant samples for HPW reconstruction. Second, reflection samples in healthy case (a) are temporally aligned ground truth ECG signal, whereas arrhythmia in case (b) disrupts this temporal consistency, causing the samples to become temporally misaligned with respect to the underlying heartbeats (details in Sec.\ref{c2}(C2)). This temporal misalignment challenges the Baseline to accurately interpret these samples for HPW reconstruction. 

Notably, our observations of these spatial disruptions are consequences of underlying altered cardiac structure \cite{prabhu2015atrial, pirruccello2024deep, li2022medical,di2012does,beinart2011left,zhang2017evaluation} and varying myocardial contraction strength \cite{butova2023inter,pidgeon1982relationship,seed1984relationships}, while the temporal disruptions are consequences of irregular cardiac contraction \cite{Vent_Dyssynchrony,solberg2020left,ciuffo2019intra,jekova2022atrioventricular,ikram1968left} commonly associated with arrhythmia, as reported in existing medical studies. These disruptions violate the core assumptions of the Baseline, which depends on locating and interpreting high-magnitude and temporally aligned reflection samples from a single, well-defined range bin for HPWs reconstruction, causing it to fail under arrhythmia. Therefore, it is imperative to develop new solutions capable of precisely locating samples of varying magnitudes across multiple affected range bins and interpreting those that are temporally misaligned to reconstruct the HPWs.

In this paper, we present \textit{mCardiacDx}, an end-to-end contactless system designed for the monitoring and diagnosis of arrhythmia. The name \textit{mCardiacDx} is derived from millimeter-wave (\underline{m}), cardiac monitoring (\underline{Cardiac}), and diagnosis (\underline{Dx}). To address the challenges, we designed mCardiacDx, which combines signal processing and deep learning to precisely locate reflection samples and reconstruct HPWs to facilitate arrhythmia monitoring and diagnosis. The architecture of mCardiacDx comprises three main components:
\begin{itemize}
    \item \textbf{Precise Target Localization (PTL)} (detailed in Section \ref{subsubsec:ptl}): To precisely locate reflection samples of varying magnitude across multiple unstable range bins, we follow a two-step process: first, we identify the target range bin, and then we refine the localization by analyzing magnitude distribution across neighboring bins.
    \item \textbf{Heart Pulse Reconstruction Network (HPR-Net.)} (detailed in Section~\ref{subsubsec:hpr}): To reconstruct HPWs, HPR-Net interprets temporally misaligned reflection samples from multiple unstable range bins leveraging PTL. It is composed of three key modules: (a) Heart Signal Extractor, (b) Encoder-Decoder, and (c) Reconstructor.
    \item \textbf{Heart Health Analysis} (detailed in Section~\ref{subsubsec:hhr}): To perform monitoring, we employ a peak detection algorithm to identify heartbeats from HPWs, calculating HR, RR intervals, and six heart rate variability (HRV) metrics. For diagnosis, we utilize a random forest model that leverages HRV metrics to diagnose arrhythmia.
\end{itemize}

We implemented mCardiacDx using Texas Instruments (TI) 's AWR1642BOOST millimeter-wave (mmWave) radar (or radar, in short) \cite{texasinstruments2024}. We evaluated mCardiacDx and PTL on a dataset of 48 subjects (24 healthy, 24 with arrhythmia) collected in real-world settings, and compared their performance against the Baseline, with all measurements referenced to ground truth ECG. Since PTL is a signal processing technique, we examined its performance by integrating it into the Baseline model, replacing the Baseline's original technique for locating reflection samples. Our results show that both mCardiacDx and PTL outperform the Baseline in monitoring and diagnosing arrhythmia. The results are detailed below:

\begin{itemize}
  \item To the best of our knowledge, mCardiacDx is the first contactless system to monitor and diagnose arrhythmia in seated conditions suitable for regular, everyday use.
   \item Both Baseline+PTL and mCardiacDx demonstrate improved HPW reconstruction. For healthy subjects, they achieved dynamic time warping (DTW) scores of 3.90 and 2.82 compared to the Baseline score of 4.02. This trend continued for arrhythmia patients, where Baseline+PTL scored 3.78, and mCardiacDx scored 2.92, significantly outperforming the Baseline score of 5.92.
   \item Both Baseline+PTL and mCardiacDx (as detailed in Section~\ref{subsec:monitoring}) outperform the Baseline in HR and RR estimation for both healthy and arrhythmia subjects. While Baseline+PTL offers notable improvements, mCardiacDx achieves the highest accuracy, especially evident in the substantial error reduction for arrhythmia patients compared to Baseline+PTL.
   \item In addition to superior monitoring performance, both Baseline+PTL and mCardiacDx achieve effective arrhythmia diagnosis performance (detailed in Section~\ref{subsubsec:diagnosis}). PTL achieved an 11\% increase in recall, a 6\% improvement in F1-score, and a 5\% rise in accuracy compared to the Baseline. mCardiacDx further enhanced these diagnostic capabilities, showing a 21\% increase in recall, a 12\% improvement in F1-score, and a 9\% rise in accuracy. Notably, both systems maintained a precision of 0.95.
\end{itemize}
\section{RELATED WORK} \label{sec:related}
In this section, we review existing approaches for monitoring heart activity, with a particular focus on arrhythmia. We categorize these approaches into two main categories: \textit{Contact-based Cardiac Sensing} and \textit{Contactless Cardiac Sensing}.

\textbf{Contact-based Cardiac Sensing:} Contact-based cardiac sensing primarily relies on two modalities: electrocardiography (ECG) and photoplethysmography (PPG). ECG has long served as the clinical gold standard for monitoring and diagnosing CVDs, such as arrhythmias and structural changes \cite{mayoclinic2024ecg}. Recent advancements have led to the development of portable ECG devices that support regular monitoring. Patch-based solutions, such as Zio Monitor \cite{yenikomshian2019cardiac} and CarePulse \cite{carepulse2025}, acquire single-lead ECG signals from the chest and have shown potential in detecting arrhythmias and premature beats using residual convolutional neural networks (CNNs) \cite{hannun2019cardiologist}. However, they all require electrodes to be attached to the skin, which may cause allergic reactions and discomfort, particularly for vulnerable populations such as newborns, the elderly, and users with skin injuries or burns. Alternatively, PPG-based wearables, such as smartwatches, provide a more convenient option by detecting blood volume changes using low-intensity light. These devices extract heart rate variability (HRV) features to support arrhythmia detection \cite{shen2019ambulatory,zhang2021towards,zhao2022robust, guo2019mobile,lopez2017oral,perez2019large}. Nonetheless, they still cause discomfort for many users, which limits their suitability for continuous and long-term monitoring \cite{jeong2017smartwatch}.

\textbf{Contactless Cardiac Sensing:}
Contactless cardiac sensing has emerged as a promising alternative to traditional contact-based methods by leveraging radio frequency (RF) signals such as WiFi \cite{zhang2017wicare,zeng2018fullbreathe}, ultra-wide-band (UWB) \cite{chen2021movi}, and mmWave radar \cite{wang2021mmhrv} to monitor heart rate (HR) and heart rate variability (HRV). These systems primarily detect thoracic wall vibrations to infer mechanical motions associated with cardiac cycles. Building on these capabilities, recent research has extended contactless sensing to support higher-level applications such as emotion recognition \cite{zhao2016emotion} and stress detection \cite{ha2021wistress}. Notably, advancements have focused on reconstructing cardiac waveforms,such as seismocardiograms (SCG), which reflect the mechanical activity of atrial and ventricular contractions. For example, radar systems enhanced with multichannel beamforming and one-dimensional (1-D) convolutional neural networks (CNNs) have been utilized to extract SCG features. \cite{ha2020contactless}. More recently, research has focused on reconstructing electrocardiogram (ECG) waveforms from RF reflections using generative models. CardiacWave \cite{xu2021cardiacwave}, for example, introduced an attention-augmented LSTM-based CaSE-ECG solver to recover ECG-like signals, while RF-ECG \cite{wang2023ecg} employed a conditional generative adversarial network (cGAN) to synthesize ECG waveforms. In parallel, other studies have explored the reconstruction of photoplethysmogram (PPG) waveforms from RF signals to infer vascular and respiratory dynamics \cite{zhang2022can,khan2022contactless}.

Distinct from prior work, \textit{mCardiacDx} and \textit{PTL} facilitate the monitoring and diagnosis of abnormal heart activity in real patients suffering from various types of arrhythmia. More specifically, we monitor and diagnose arrhythmia by reconstructing heartbeat waveforms from spatially and temporally disrupted reflections. This capability aligns with established clinical findings and demonstrates the robustness of \textit{mCardiacDx} and \textit{PTL} in capturing abnormal cardiac patterns across diverse arrhythmic conditions.
\section{Designing mCardiacDx}
\label{sec:sysdesi}
We design mCardiacDx through two distinct stages. First, we present challenges in reconstructing HPWs in Section~\ref{subsec:challenges}. Then, we construct mCardiacDx in Section~\ref{subsec:mcardiacDX}.

\subsection{Challenges in Heart Pulse Waveform Reconstruction} \label{subsec:challenges}

\subsubsection{Radar-Based Cardiac Sensing} \label{subsubsec:rf_sensing}
The human heart consists of four chambers—the left atrium (LA), right atrium (RA), left ventricle (LV), and right ventricle (RV)—that manage blood circulation through synchronous contraction. The atria receive blood and pump it into the ventricles, which then store and propel blood throughout the body. The mechanical activity of the heart is generated under the stimulation of electrical activity; i.e., mechanical contraction of the heart is initiated by electrical depolarization \cite{pfeiffer2014biomechanics}. This activity is regulated by the heart’s electrical conduction system, which initiates at the sinoatrial (SA) node and propagates through the atrioventricular (AV) node, right bundle branch (RBB), left bundle branch (LBB) and the His–Purkinje fibers (PF) network. In healthy subjects, the atria and ventricles maintain structural stability and contract synchronously \cite{ripa2023physiology,HeartChambers,bowman2003assessment,maeda2003quantification,NHLBIanatomy}. However, this structural stability \cite{prabhu2015atrial, pirruccello2024deep, li2022medical,di2012does,beinart2011left,zhang2017evaluation} and synchronous contraction are significantly disrupted in the case of arrhythmia. 

\begin{figure}[h]
  \centering
  \includegraphics[width=0.94\columnwidth, height=0.48\columnwidth]{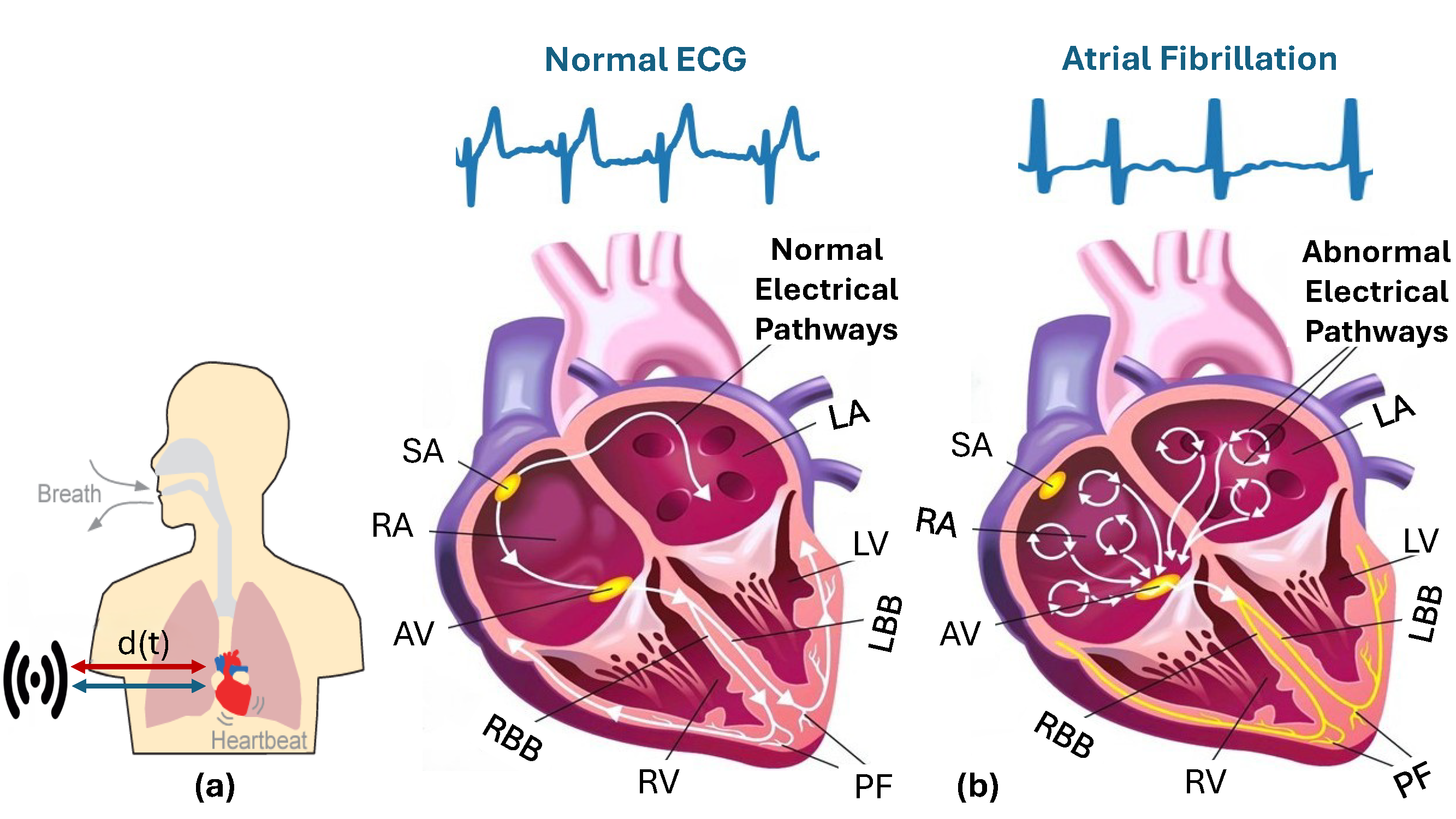} %
  \caption{(a) Radar-based sensing of cardiac activity. (b) Structural and electrical conduction pathways of the heart: the left panel shows a healthy heart with a normal ECG, while the right panel depicts an arrhythmic heart with an abnormal ECG (atrial fibrillation).}
  \label{fig:heart_normal_afib}
  \Description{}
\end{figure}
Figure~\ref{fig:heart_normal_afib} illustrates (a) radar-based cardiac sensing and (b) the structure and electrical pathways of a healthy heart (left) and one with arrhythmia (right). During each heartbeat, mechanical contraction causes subtle vibrations at different regions on the chest wall \cite{he2009learning,ning2021rf}, which are captured by a radar. The radar transmits mmWaves to the chest wall and receives the reflected signals (or simply `reflection samples') that contains cardiac vibration information. These reflections are represented by the 2-D channel impulse response (CIR) matrix, with its dimensions denoted as fast time and slow time due to their different sampling rates \cite{zheng2021enhancing,chen2021movi}. Fast-time samples indicate the range of the reflecting surface, defining range bins where each bin corresponds to a specific spatial region on the chest wall. Slow-time is fast-time snapshots of reflection samples at a much lower rate over repeated transmissions, capturing the temporal evolution of samples within each range bin. Baseline analyzes these reflection samples across the fast-time (range) and slow-time (temporal) dimensions to extract the chest wall vibration, which is then transformed into HPWs for HR and RR interval estimation. 

We define the chest wall vibration at time \textit{t} as $d(t)$, which can be extracted from the phase $\phi(t)$ of the reflection samples using Equation~\ref{eq:phase}, where $\lambda$ is the mmWave wavelength.

\begin{equation}
    d(t) \propto \angle \text{CIR}(t) = \frac{\lambda}{4\pi} \phi(t) \quad
    \label{eq:phase}
\end{equation}
However, accurately analyzing these reflection samples to extract chest wall vibration from the phase poses a significant challenge in arrhythmia patients. This challenge stems from the disrupted structural stability and asynchronous contraction of the heart, leading to profound alterations in chest wall vibration patterns. These alterations directly manifest as a disruption in the spatial stability and temporal consistency of the reflection samples in the range bins. As a result, it becomes very challenging to locate and accurately interpret these samples for robust HPW reconstruction, ultimately impacting HR and RR interval estimation. We outline these challenges in \textbf{(C1)} and \textbf{(C2)} below.

\begin{figure}[h]
  \centering
  \begin{tabular}{cc}
    \includegraphics[width=0.46\linewidth, height=0.36\linewidth]{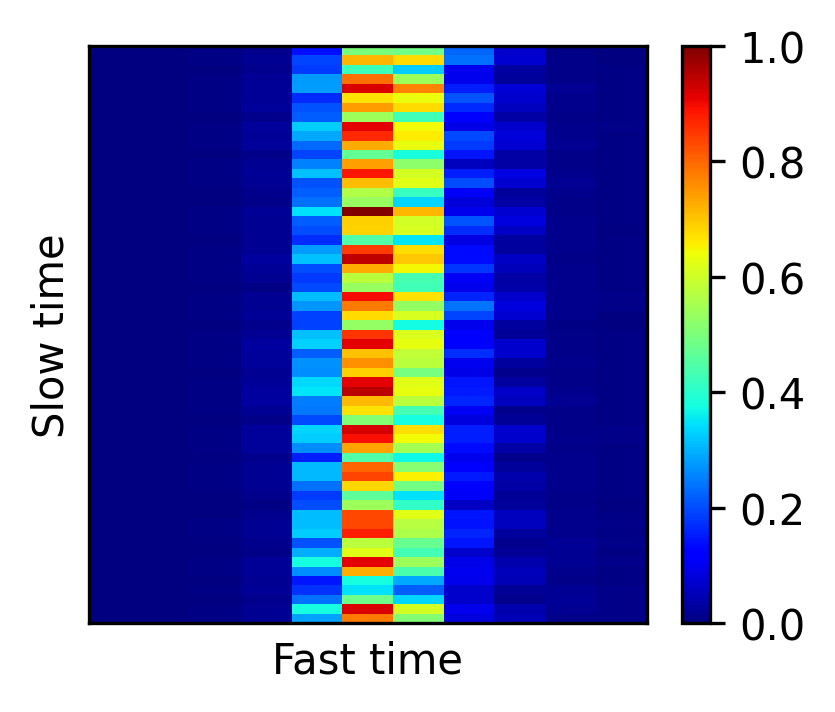} & 
    \includegraphics[width=0.46\linewidth, height=0.36\linewidth]{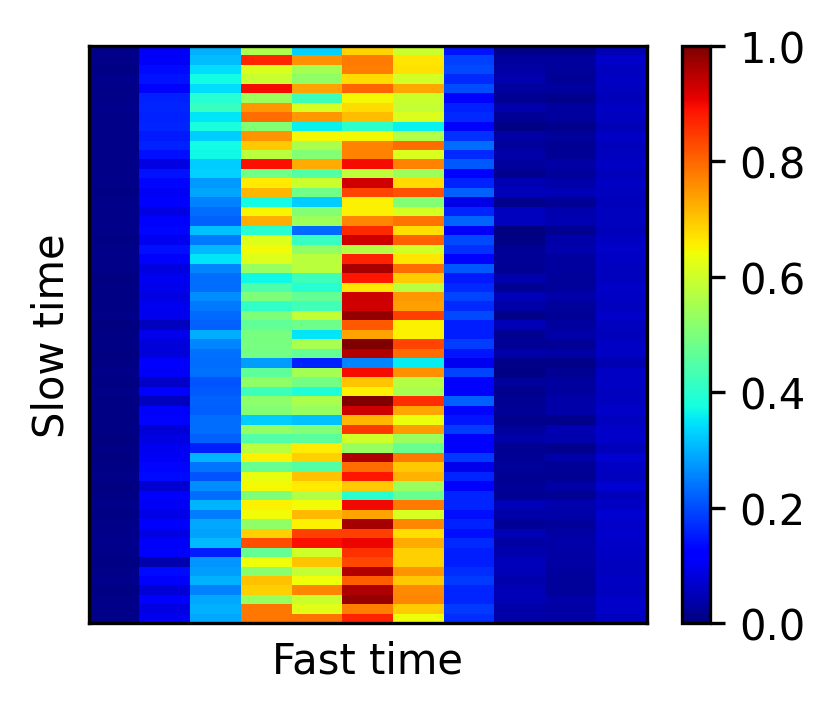} \\ 
    (a) Healthy Subject & (b) Arrhythmia Patient \\
  \end{tabular}
  \caption{Heart reflection patterns in healthy subjects and arrhythmia patients.}
  \label{fig:cir_mag_both}
  \Description{}
\end{figure}

\textit{\textbf{C1: Precisely Locating Dispersed Reflection Samples from Multiple Unstable Range Bins.}}\label{c1} 
We observe that high-magnitude reflection samples are predominantly concentrated within a single range bin in the fast-time dimension for healthy subjects (Figure~\ref{fig:cir_mag_both}(a)). This spatial concentration indicates that these reflection samples correspond to the chest wall vibration at a single spatially stable region. Their consistently high magnitude (represented by the reflection sample intensity) over slow time suggests strong vibration of the region. We attribute these observed concentration and magnitude as direct indicators of underlying structurally intact cardiac chambers and their strong contraction strength common in healthy subjects \cite{ripa2023physiology,HeartChambers,bowman2003assessment}. Lower-magnitude samples also appear in neighboring bins, but they are not useful for further analysis. The Baseline, designed to locate high-magnitude reflection samples from a single stable range bin, successfully identifies these reflection samples, which allows the reconstruction of HPWs for estimating HR and RR intervals in healthy subjects.

In contrast, we observe that this spatial stability is disrupted in arrhythmia patients (Figure~\ref{fig:cir_mag_both}(b)). Reflection samples are no longer confined to a single range bin but are instead dispersed across multiple unstable range bins along the fast-time dimension. This spatial dispersion suggests that these reflection samples correspond to chest wall vibration at multiple unstable regions. Concurrently, their varying magnitudes over slow time indicate varying vibrational strength of each region. We interpret these observed disruption across multiple unstable range bins and varying magnitude as direct indicators of the underlying structurally altered cardiac chambers \cite{prabhu2015atrial, pirruccello2024deep, li2022medical,di2012does,beinart2011left,zhang2017evaluation} and their varying contraction strength \cite{butova2023inter,pidgeon1982relationship,seed1984relationships}, both common in arrhythmia patients as per medical studies findings. This complex pattern across multiple unstable range bins with varying magnitudes challenges the Baseline. As a result, the Baseline struggles to precisely locate these diverse reflection samples, which hinders HPW reconstruction for HR and RR interval estimation in arrhythmia patients. 

\textit{\textbf{C2: Interpreting Temporally Misaligned Reflection Samples from Multiple Unstable Range Bins.}}  \label{c2} While \textbf{C1} addressed the challenge of precisely locating dispersed reflection samples, \textbf{C2} focuses on the equally critical problem of interpreting their temporal consistency for robust HPW reconstruction. We investigate the temporal alignment of reflection samples corresponding to different vibrating regions on the chest wall. Temporal alignment is critical for achieving consistent phase across reflection samples, a prerequisite for accurate HPW reconstruction. Misaligned reflection samples introduce phase inconsistencies that complicate interpretation for HPW reconstruction.

\begin{figure}[h]
  \centering
  \begin{tabular}{cc}
    \includegraphics[width=0.46\columnwidth, height=0.40\columnwidth]{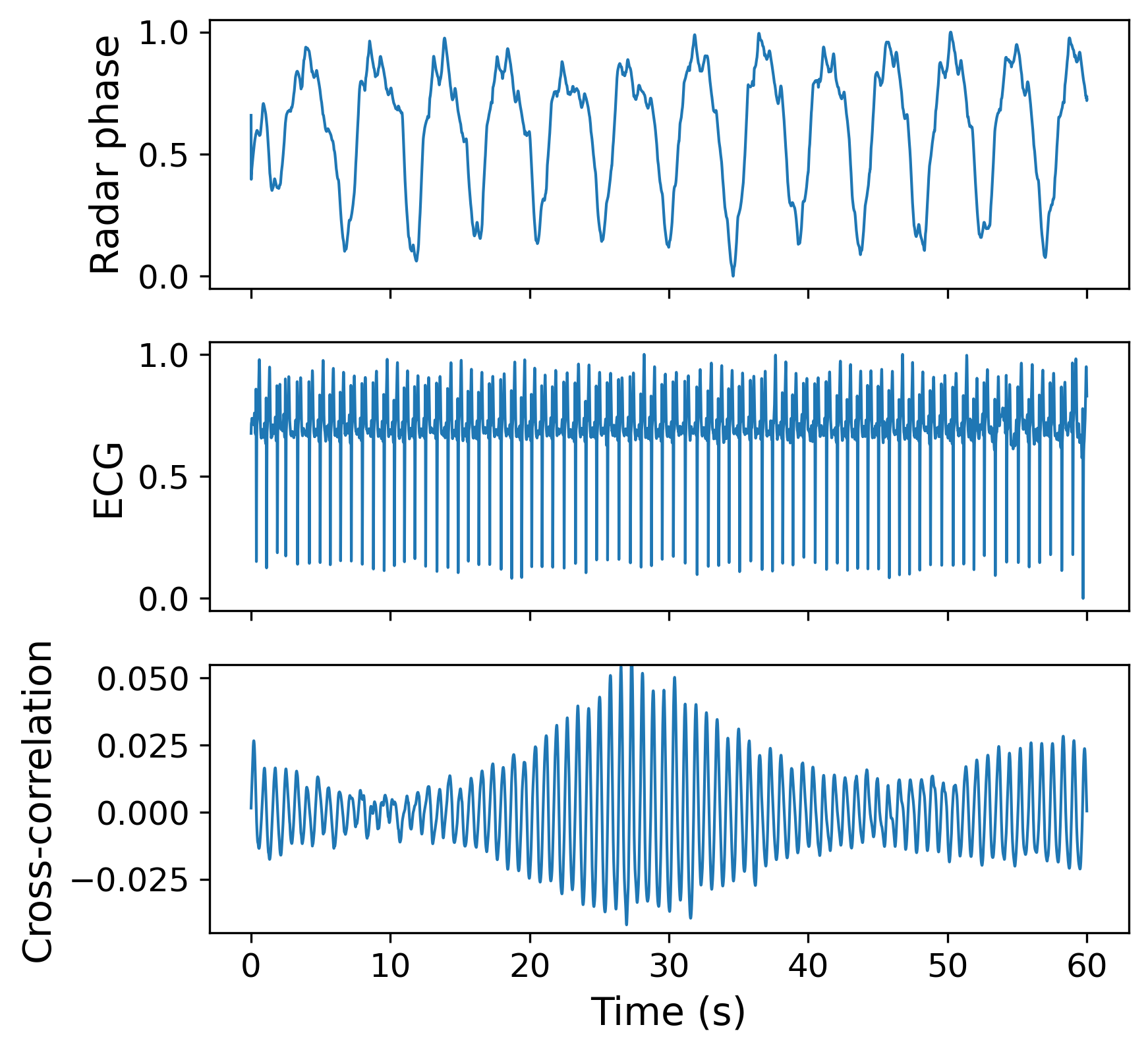} & 
    \includegraphics[width=0.46\columnwidth, height=0.40\columnwidth]{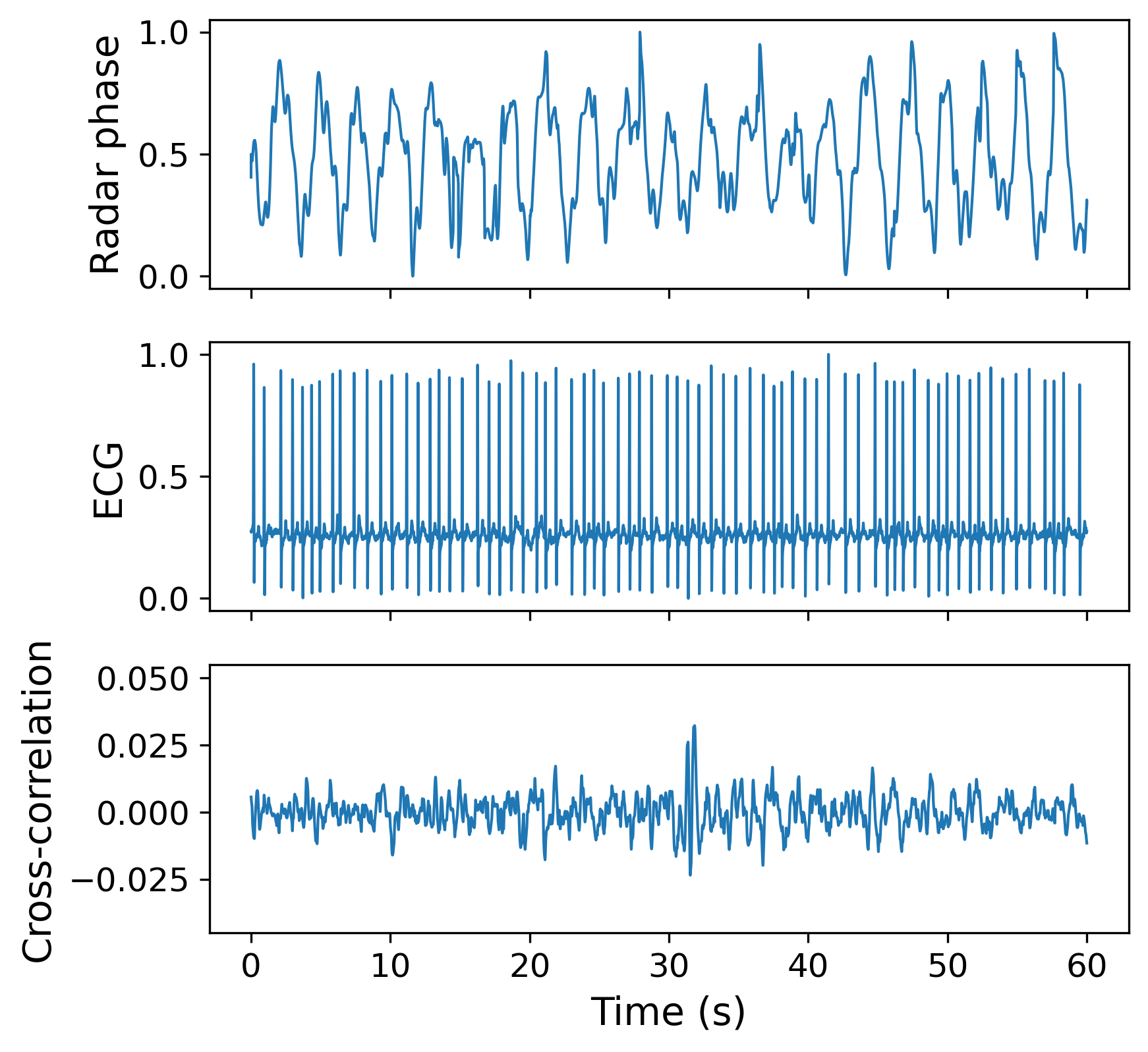} \\ 
    (a) Healthy Subject & (b) Arrhythmia Patient \\
  \end{tabular}
  \caption{Cross-Correlation between the radar phase and ECG signals. }
  \label{fig:cc_both}
  \Description{}
\end{figure}

To assess this, we analyze the temporal alignment between the radar phase extracted from reflection samples at different range bins and the ECG signal. The radar phase captures the timing of chest wall vibrations at different range bins, while the ECG provides a reference for the underlying heart contractions, represented by the R peaks that generate this vibration. We compute the cross-correlation (CC) between the radar phase and ECG signal to quantify their temporal alignment as a function of time lag. The Zero-Normalized Cross-Correlation coefficient (ZNCC) is used as the alignment metric (see Figure~\ref{fig:cc_both}). For relative comparison, we normalize ZNCC values by those observed in healthy subjects.

We observe strong temporal alignment between the radar phase and ECG signal for healthy subjects, with a normalized ZNCC of 1.0. As illustrated in Figure~\ref{fig:cc_both}(a), the cross-correlation shows consistent peaks that reliably align with the ECG-defined R peaks. This consistency confirms that the radar phase is temporally aligned with the R peaks and thus the corresponding reflection samples are also temporally aligned. This alignment indicates that the underlying contractions occur in sync with the ECG-defined R peaks, which is consistent with established clinical findings \cite{maeda2003quantification,NHLBIanatomy}, thereby further validating our observations. Such aligned reflection samples from a single stable range bin allow the Baseline to accurately interpret samples for HPW reconstruction, which facilitates HR and RR interval estimation in healthy subjects.

Conversely, arrhythmia patients show significantly lower temporal alignment between the radar phase and ECG signal, with a normalized ZNCC of 0.54. The cross-correlation results shown in Figure~\ref{fig:cc_both}(b) reveal inconsistent peaks that do not align with the ECG-defined R peaks, indicating a less stable temporal relationship. This inconsistency suggests that the radar phase is not temporally aligned with the R peaks and thus the corresponding reflection samples are also not temporally aligned. This lack of alignment indicates that underlying contractions vary with respect to the R peaks of the ECG. Such variability aligns with established clinical findings that demonstrate out-of-sync or asynchronous heart contractions during arrhythmias \cite{Vent_Dyssynchrony,solberg2020left,ciuffo2019intra,jekova2022atrioventricular,ikram1968left}, further validating our observations. The presence of misaligned reflection samples from multiple unstable range bins poses a challenge to the Baseline, which is designed to interpret only temporally aligned reflection samples. As a result, it fails to accurately interpret these reflection samples, which hinder HPW reconstruction for HR and RR interval estimation in arrhythmia patients.

To address the challenges of precise localization (C1) and accurate interpretation of temporally misaligned (C2) reflection samples from multiple unstable range bins, we developed a novel signal processing technique, Precise Target Localization (PTL), and a novel deep learning network, Heart Pulse Reconstruction Network (HPR-Net). We discuss these implementation details in the subsequent sections.

\subsection{Constructing mCardiacDx} \label{subsec:mcardiacDX}
In this section, we detail the construction of mCardiacDx with the following key components:
\begin{itemize}
    \item \textbf{Precise Target Localization (PTL):} An advanced signal processing technique designed to precisely locate reflection samples from multiple unstable range bins as well as from single range bin using a dynamic processing strategy, addressing (C1).
    \item \textbf{ Signal Processing:} Extracts cardiac motion information from PTL's selected reflection samples using bandpass and second-order derivative filters to suppress interference and emphasize cardiac signals.
    \item \textbf{Heart Pulse Reconstruction Network (HPR-Net.):} A deep learning network that uses PTL to locate and interpret reflection samples and reconstructs HPWs as Gaussian pulse trains, addressing (C2).
    \item \textbf{Heart Health Analysis:} Monitors HR and RR intervals based on HPWs and diagnoses arrhythmia based on HR, RR intervals, and other HRV metrics.
\end{itemize}
\begin{figure*}[h]
    \centering
    \includegraphics[width=\linewidth]{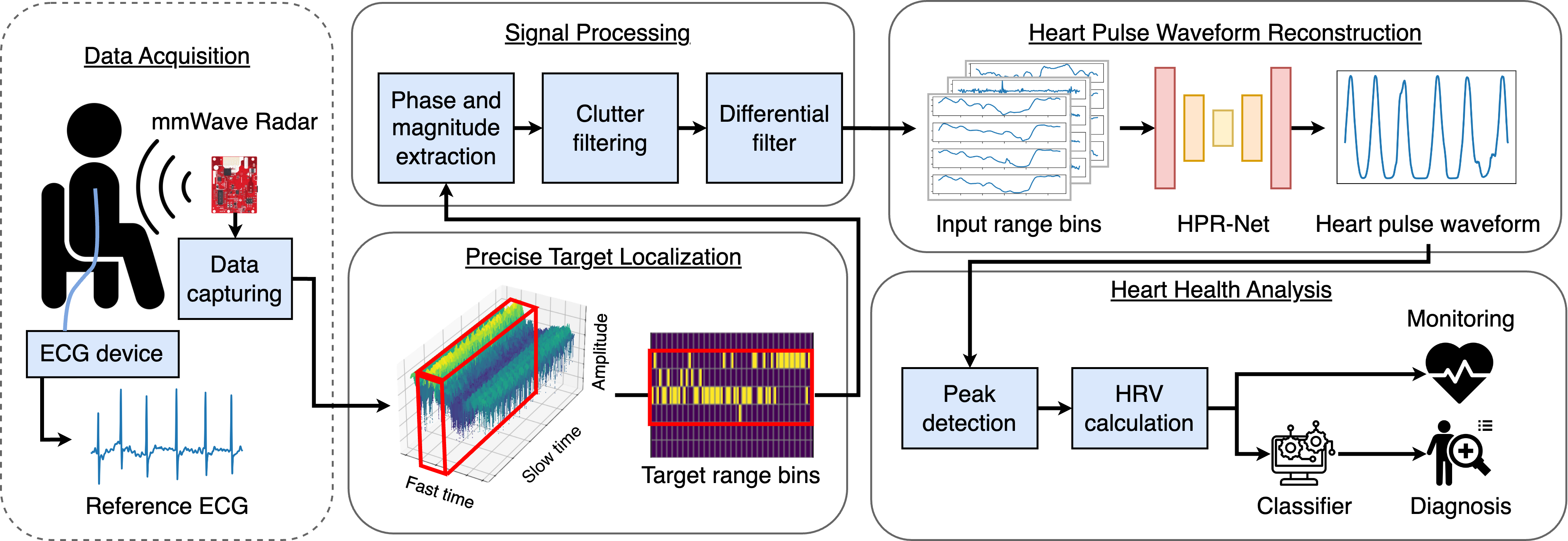}
    \caption{Schematic overview of mCardiacDx}
    \label{fig:mcardiacdx}
    \Description{}
\end{figure*}

\subsubsection{Precise Target Localization}\label{subsubsec:ptl}
Accurate localization of varying magnitude reflection samples from multiple unstable range bins is crucial for accurate HPW reconstruction and low estimation errors in HR and RR interval estimation for arrhythmia monitoring and diagnosis. The PTL algorithm operates through a sequential two-step process, as detailed in Algorithm~\ref{alg:ptl}. In the initial step, the algorithm identifies the target range bin corresponding to the highest magnitude reflection samples. It does this by detecting the high magnitude samples across chirps using the Most Common Bin (MCB) algorithm \cite{assana2020cardiovascular,ha2020contactless} (lines 1-2). This process enables PTL to locate high magnitude reflection samples consistently concentrated in a single stable range bin. However, this initial step is inadequate for precisely locating reflection samples of varying magnitudes dispersed across multiple unstable range bins, as such samples may dynamically shift between different range bins over time.

To address this limitation, PTL enhances the target selection by implementing a dynamic processing strategy in the second step. It first defines a neighboring window around the target range bin $t$ by computing the first and last bins based on a predefined window breadth $w_b$ (line 3). The $w_b$ is a parameter that determines how many adjacent range bins are considered. This step extracts a sub-matrix that includes surrounding range bins (line 4), enabling the algorithm to capture reflection samples of varying magnitude that dynamically shift across neighboring range bins over time, thereby addressing the limitation of the initial step.

However, merely expanding the range bin selection does not fully account for the dynamic spatial shifts of these reflection samples over time. As these shifts can evolve, it is crucial to continually monitor and track them for precise localization. To address this, PTL iteratively processes chirps within a sliding time window (line 7),  extracts data for the current window (line 8), reapplies the MCB algorithm to refine the target range bin $t$ (line 9), and tracks these spatial variations in the identified range bin over time (lines 9–10). This iterative adjustment allows PTL to successfully locate reflection samples dispersed across multiple unstable range bins with varying magnitude, potentially linked to health conditions of structurally altered chambers and their varying contraction strength as mentioned in C1. The details of the PTL algorithm are presented in Algorithm~\ref{alg:ptl}.

\begin{figure}[h]
  \centering
  \begin{tabular}{cc}
    \includegraphics[width=0.46\linewidth, height=0.36\linewidth]{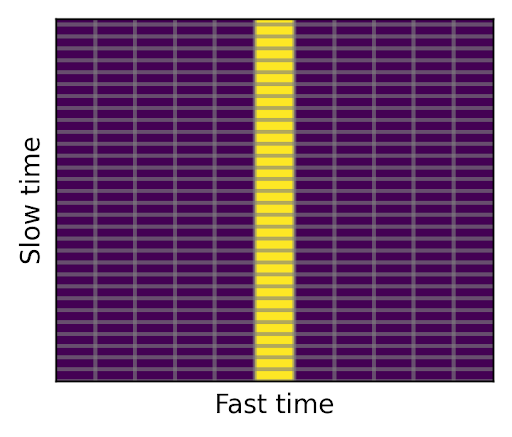} & 
    \includegraphics[width=0.46\linewidth, height=0.36\linewidth]{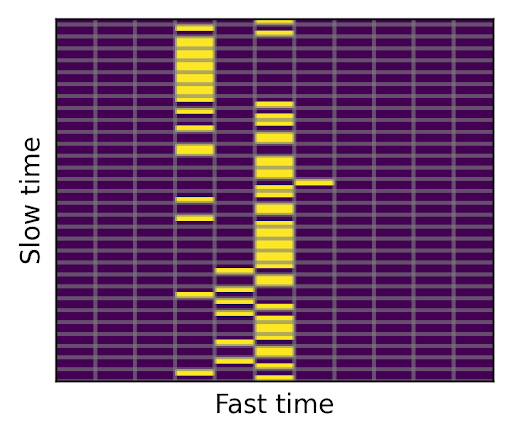} \\ 
    (a) Healthy Subject & (b) Arrhythmia Patient \\
  \end{tabular}
  \caption{Reflections captured by PTL for healthy subject and patient with arrhythmia}
  \label{fig:cir_ptl_both}
  \Description{}
\end{figure}

We validated the ability of PTL to locate reflection samples concentrated in a single, stable range bin, and multiple unstable range bins. As shown in Figure ~\ref{fig:cir_ptl_both}, PTL accurately identified high magnitude reflection samples consistently appearing in a single, stable range bin and successfully located the samples of varying magnitude dispersed across multiple unstable range bins for respective subjects in Figure~\ref{fig:cir_ptl_both}(a) and (b). This groundbreaking algorithm outperforms conventional baseline algorithms by improving HPW reconstruction, reducing errors in HR and RR interval estimation for monitoring, and advancing the diagnostic assessment of arrhythmia, as detailed in Section \ref{sec:perf}.
\begin{algorithm} 
\caption{PTL Algorithm}\label{alg:ptl}
\begin{algorithmic}[1]
\Require CIR matrix $R \in \mathbb{C}^{\text{num\_samples} \times \text{num\_chirps}}$,\newline Time window size (in number of chirps) $w_t \in \mathbb{Z}$,\newline Range bin window size $w_b \in \mathbb{Z}$
\Ensure Selected range bins at each chirp $s \in \left[1, \text{num\_samples}\right]^{\text{num\_chirps}}$
\State $M \gets \abs{\left(R\right)}$    
\State $t \gets \Call{MostCommonBin}{M}$    
\State $\text{first\_bin} \gets t - \lfloor \frac{w_b}{2} \rfloor$, $\text{last\_bin} \gets t + \lceil \frac{w_b}{2} \rceil - 1$
\State $M \gets M[\text{first\_bin}: \text{last\_bin}, :]$  
\State $i \gets 1$
\State $s \gets \mathbf{0}^{\text{num\_chirps}}$   
\While{$i \leq \text{num\_chirps}$}
    \State $M' \gets M\left[:, i: i + w_t - 1\right]$   
    \State $s_{i:i + w_t - 1} \gets \text{first\_bin} + \Call{MostCommonBin}{M'}$   
    \State $i \gets i + w_t$
\EndWhile
\end{algorithmic}
\end{algorithm}

\subsubsection{Signal Processing}\label{subsubsec:sp}
This component transforms the reflection samples selected by PTL into cardiac motion features suitable for HPW reconstruction. We first extract the phase and magnitude signals from the reflection samples, which represent the underlying chest wall vibration and its strength, respectively. However, these initial signals contain various physiological motions, including respiration, heartbeat, and other body motions. To isolate the relevant cardiac motion, we apply a bandpass filter (0.2–50 Hz) as a clutter filter. This removes low-frequency baseline drift and high-frequency noise, retaining primarily respiration and cardiac motions. While this filters out general noise, isolating cardiac motion from respiration remains critical. Chest wall vibration caused by respiration is slow with low acceleration, whereas cardiac activity, such as heartbeats, induces significant acceleration. We leverage this fundamental difference in acceleration profiles to distinguish cardiac motion. Therefore, we compute the chest acceleration as the second-order derivative \cite{Holoborodko2024} of the chest vibration, which serves as a refined cardiac motion feature that selectively enhances the rapid, high-acceleration components indicative of heartbeats. This filter is defined by Equation~\ref{eq:differential-filter}, where $\phi_t$ represents the phase signal at time $t$, and $h$ denotes the sampling interval. Finally, this component provides three distinct processed output, the bandpass-filtered phase and magnitude signals, and the second-order derivative-filtered phase signal—for the subsequent HPR-Net.
\begin{equation}
    \phi''_t = \frac{(\phi_{t-3} + \phi_{t+3}) + 2(\phi_{t-2} + \phi_{t+2}) - (\phi_{t-1} + \phi_{t+1}) - 4\phi_t}{16h^2}
    \label{eq:differential-filter}
\end{equation}

\subsubsection{Heart Pulse Reconstruction Network (HPR-Net)} \label{subsubsec:hpr}
\begin{figure}[H]
    \centering
    \includegraphics[width=0.95\linewidth]{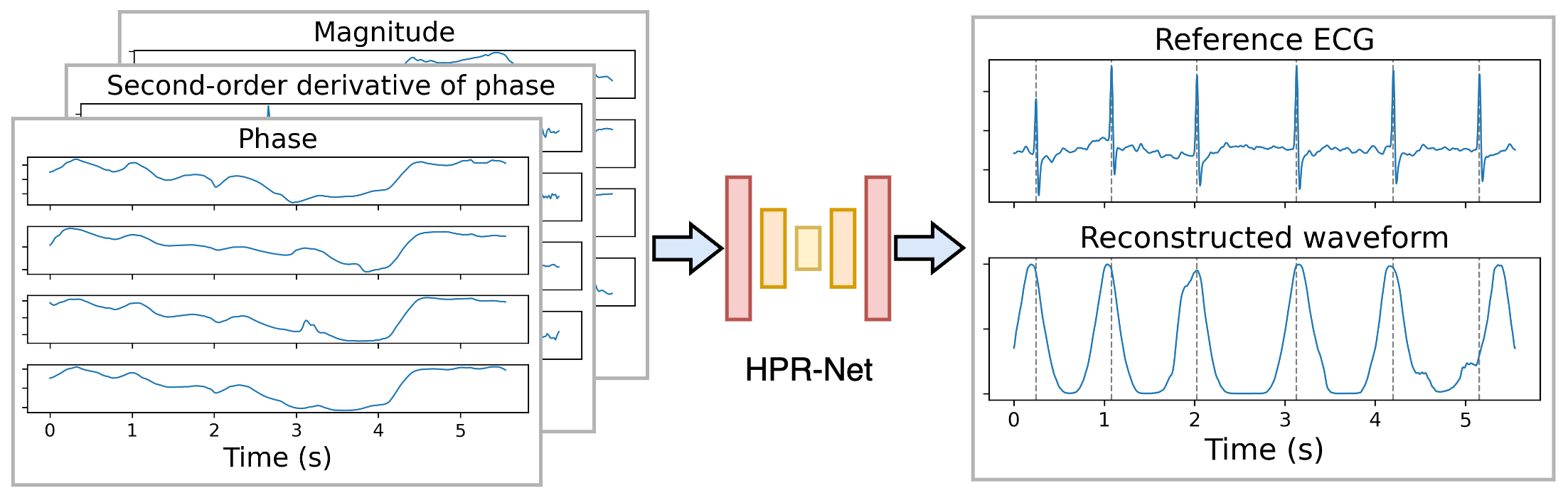}
    \caption{Example input and output of HPR-Net.}
    \label{fig:hprnet_input_output}
    \Description{}
\end{figure}

\begin{figure}[h]
    \centering
    \includegraphics[width=1.0\linewidth]{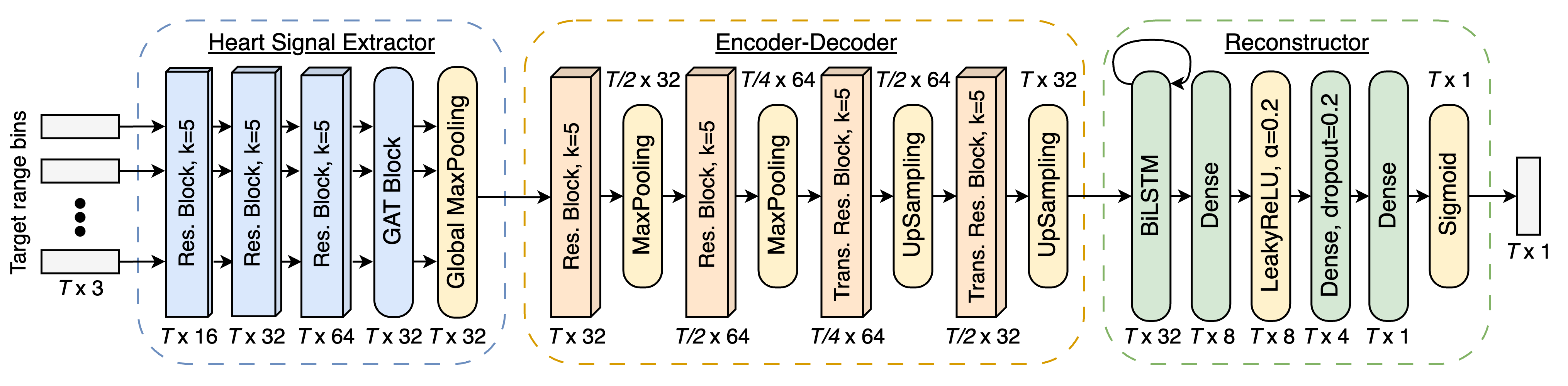}
    \caption{Architectural overview of the proposed Heart Pulse Reconstruction Network (HPR-Net). The kernel size of each residual block is denoted by k, and each layer is annotated with the shape of its output.}
    \label{fig:hprnet_architecture}
    \Description{}
\end{figure}

HPR-Net reconstructs HPWs as a series of Gaussian pulse trains. Its end-to-end neural network utilizes phase, the second-order derivative of phase, and magnitude as inputs to extract cardiac motion features for reconstructing HPWs, as shown in Figure~\ref{fig:hprnet_input_output}. Unlike the Baseline that interprets the reflection samples to extract cardiac motion features from temporally aligned samples concentrated to a single stable range bin, HPR-Net interprets the samples, extracting these features from temporally misaligned samples dispersed across multiple unstable range bins by leveraging the PTL algorithm (\ref{alg:ptl}). To select the inputs from these multiple range bins, we employ an attention mechanism inspired by graph attention networks (GAT) \cite{velickovic2017graph}. This mechanism captures inter-bin correlations at each timestep to enable the network to learn inter-bin cardiac motion features. These features are then used to derive a cardiac motion latent representation that ultimately leads to the construction of HPWs. The architectural framework of HPR-Net comprises three distinct modules: (1) Heart Signal Extractor, (2) Encoder-Decoder, and (3) Reconstructor, as depicted in Figure~\ref{fig:hprnet_architecture}. Below are detailed descriptions of each module.

\begin{figure}[h]
    \centering
    \includegraphics[width=0.5\linewidth]{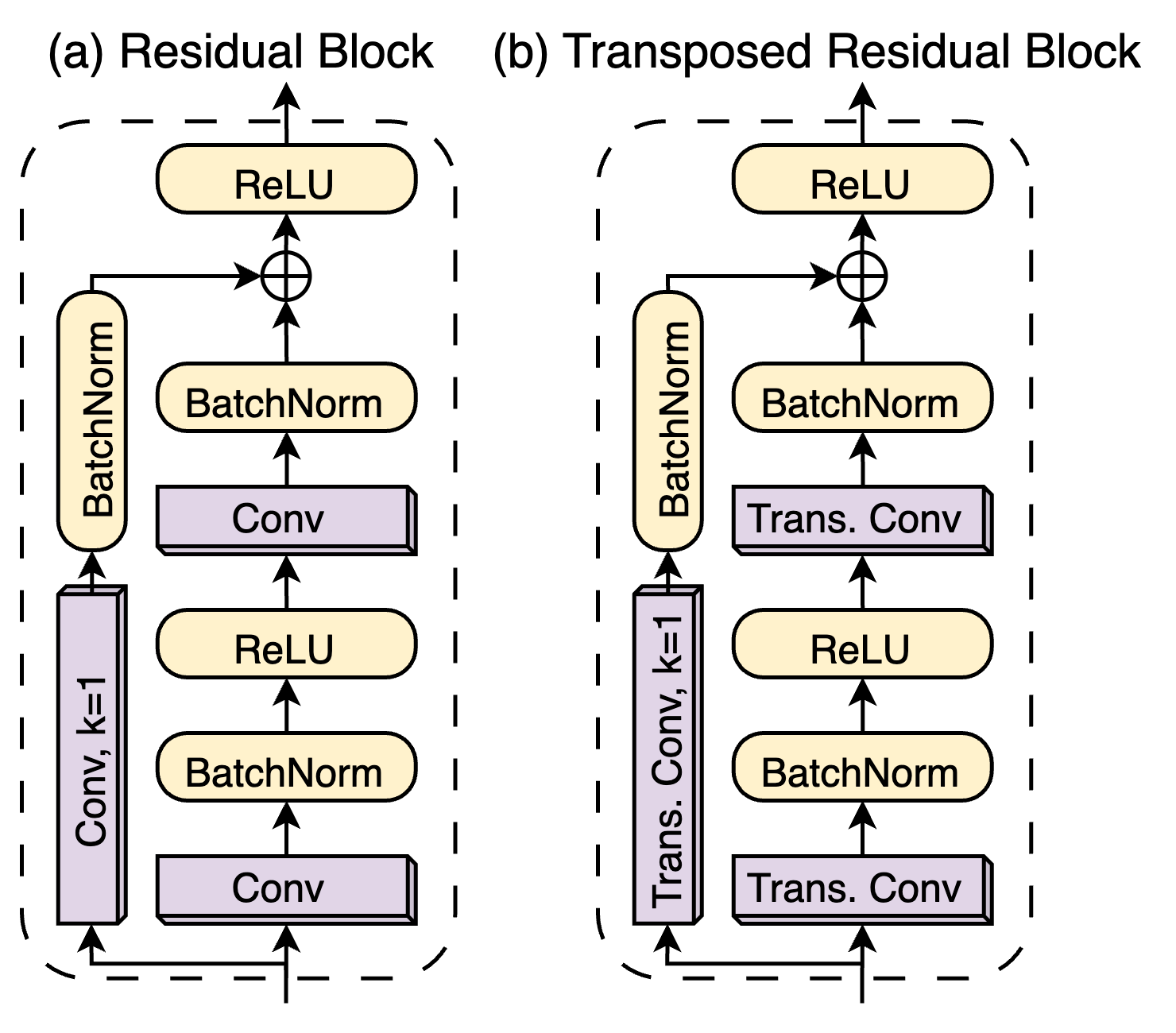}
    \caption{HPR-Net building blocks: (a) residual block and (b) transposed residual block.}
    \label{fig:hprnet_building_blocks}
    \Description{}
\end{figure}

\textbf{(1) Heart Signal Extractor:} 
This module is designed to extract inter-bin cardiac motion features from the inputs. We begin employing one-dimensional convolutional operations to capture the temporal information within each range bin. To refine the extraction process, we utilize a variant of residual blocks, as depicted in Figure~\ref{fig:hprnet_architecture}. These residual blocks are applied across all range bins selected through PTL, refining the initial features and improving performance and learning efficiency \cite{He16}.

Subsequently, a Graph Attention Network (GAT) mechanism \cite{Veli18} is implemented to model correlations between the range bins. Unlike traditional convolutional networks \cite{Chang22,Xu23}, which assume local spatial correlations, GAT effectively detects relationships between non-adjacent range bins.This capability allows for the accurate detection of patterns within the input features that indicate cardiac motion, thereby facilitating the extraction of relevant inter-bin cardiac motion features. Specifically, at each time step, the GAT block processes a set of temporal features from the range bins, denoted as $\vec{h_i} \in \mathbb{R}^d$ for $i=1,2,...,m$, where $d$ is the dimension of the input features resulting from the convolutional operations and $m$ represents the number of range bins. The learned representation of each range bin's features, $\vec{h'_i} \in \mathbb{R}^{d'}$, where $d'$ is the length of the output features, is computed as:
\begin{displaymath} \vec{h_i'}=\sigma\left(\sum_{j=1}^{m}{\alpha_{ij}\boldsymbol{W}\vec{h_j}}\right) \end{displaymath} 

Here, $\boldsymbol{W} \in \mathbb{R}^{d' \times d}$ is a learnable weight matrix, and $\sigma\left(\cdot\right)$ denotes the activation function. We employ the Leaky ReLU function with a negative slope of 0.2 as the activation function. The attention coefficient $\alpha_{ij}$ of $\vec{h_j}$ with respect to $\vec{h_i}$ is calculated by:
\begin{displaymath} \alpha_{ij} = \frac{\exp\left(\text{LeakyReLU}\left(\vec{a}^T\left[\boldsymbol{W}\vec{h_i} || \boldsymbol{W}\vec{h_j}\right]\right)\right)}{\sum_{k=1}^{m}{\exp\left(\text{LeakyReLU}\left(\vec{a}^T\left[\boldsymbol{W}\vec{h_i} || \boldsymbol{W}\vec{h_k}\right]\right)\right)}} \end{displaymath}

where $\vec{a} \in \mathbb{R}^{2d'}$ is a learnable weight vector, $\left[\cdot||\cdot\right]$ denotes vector concatenation, and $\text{LeakyReLU}(\cdot)$ specifies the element-wise Leaky ReLU function with a negative slope of 0.2. The GAT block outputs the inter-bin cardiac motion features within these range bin representations, which are then used by the encoder.

\textbf{(2) Encoder-Decoder:} 
The encoder module employs a convolutional encoder-decoder (CED) network to construct a cardiac motion latent representation from the extracted inter-bin cardiac motion features. CED networks are known for their efficacy in time series analysis, anomaly detection, and the compression and reconstruction of physiological data, such as ECG \cite{cednet,fotiadou2020end,llugsi2021novel}. These networks learn an efficient latent representation of data by transforming the input into a lower-dimensional latent space and then decoding this compressed representation to recover the original data.

Our approach utilizes a CED network with skip connections to achieve this task. The encoder component of the network employs a stack of residual blocks to capture temporal information from the inter-bin cardiac motion features. These residual blocks facilitate the extraction of pertinent temporal characteristics. In contrast, the decoder component employs transposed residual blocks, as shown in Figure~\ref{fig:hprnet_building_blocks}, to reconstruct the cardiac motion latent representation from the encoded latent space. This design enables the effective formation of a high-level cardiac motion latent representation. 

\textbf{(3) Reconstructor:} 
The reconstructor module employs a Bidirectional Long Short-Term Memory (BiLSTM) network to generate the reconstructed heart pulse waveform from the cardiac motion latent representation. Long Short-Term Memory (LSTM) networks are well-suited for modeling long-range dependencies in time series data, and their bidirectional variant (BiLSTM) further enhances performance by processing the input sequence in both forward and backward directions \cite{Graves05,Hochreiter97}.

In our approach, the cardiac motion latent representation produced by the encoder-decoder module at each time step is fed into the BiLSTM network, as depicted in Figure~\ref{fig:hprnet_architecture}. The BiLSTM captures temporal dependencies and provides a comprehensive understanding of the cardiac motion dynamics, capturing both short-term and long-term relationships in cardiac motion. Following this, a series of dense layers and a sigmoid activation function are applied to compute the amplitude of the cardiac motion waveform at each time step, thereby generating the reconstructed waveform (HPWs).

\subsubsection{Heart Health Analysis} \label{subsubsec:hhr}
This component performs monitoring and diagnosis based on the reconstructed HPWs.  For monitoring, we use a peak detection algorithm \cite{} to identify the peaks of the HPWs and estimate HR and RR intervals.  Subsequently,  we perform a Heart Rate Variability (HRV) analysis. Our focus is on six widely accepted HRV metrics, as they serve as effective indicators of cardiac autonomic function.  Studies often link changes in these metrics to various cardiac arrhythmias \cite{huikuri2013heart} and  autonomic stress responses \cite{ma2022longitudinal}.  These metrics include the mean of RR intervals (MeanNN), the median of RR intervals (MedianNN), the standard deviation of RR intervals (SDNN), the interquartile range of RR intervals (IQRNN), the median absolute deviation of RR intervals (MadNN), and the ratio of MadNN to MedianNN, all as defined in \cite{neurokit2, neurokit2_hrv, shaffer2017overview}.

For diagnosis, we employ a Random Forest classifier \cite{breiman2001random,yang2021ensemble}. This classifier takes HRV metrics derived from both the actual ECG readings and the reconstructed HPWs as its input feature set to diagnose arrhythmias. This approach enables the classifier to utilize information on cardiac variability from both real cardiac activity (ECG) and simulated cardiac activity (HPWs), enhancing its ability to differentiate between healthy and arrhythmic conditions. We detail the training and performance of monitoring and diagnosis in Sections \ref{sec:training} and \ref{sec:perf}.

\section{IMPLEMENTATION AND EXPERIMENTAL DETAILS} \label{sec:implen}

\textbf{Hardware and Software Toolkits:} We used a texas instruments millimeter wave board (AWR1642BOOST) with the DCA1000 real-time data-capture adapter for precise radar sensing \cite{texasinstruments2024,ti2024}. The board operates in the 77–81 GHz range. We configured the system with 1 transmitter (1Tx) and 4 receivers (4Rx) to enhance spatial resolution while maintaining high temporal resolution through a short chirp period (50 µs) and a high sampling rate (6000 ksps). Each chirp was followed by an idle time of 150 µs, with 256 samples captured per chirp. The receive gain was set to 30 dB, and data acquisition was managed through mmWave Studio \cite{ti2024mmwave}, which streamed the data to a host PC for processing. Simultaneously, we recorded ECG signals using the shimmer 3TM ECG development kit \cite{shimmer3_ecg} at a sampling rate of 500 Hz. Radar signal processing was implemented using C/C++, while neural network components were developed in python with tensorflow 2.10 \cite{tensorflow} on a workstation with an Intel i7 CPU, 32GB DDR4 RAM, and a NVIDIA GeForce RTX 2070 graphics card \cite{nvidia2025geforce20}.

\begin{figure}[h]
  \centering
  \begin{tabular}{cc}
    \includegraphics[width=0.46\linewidth, height=0.36\columnwidth]{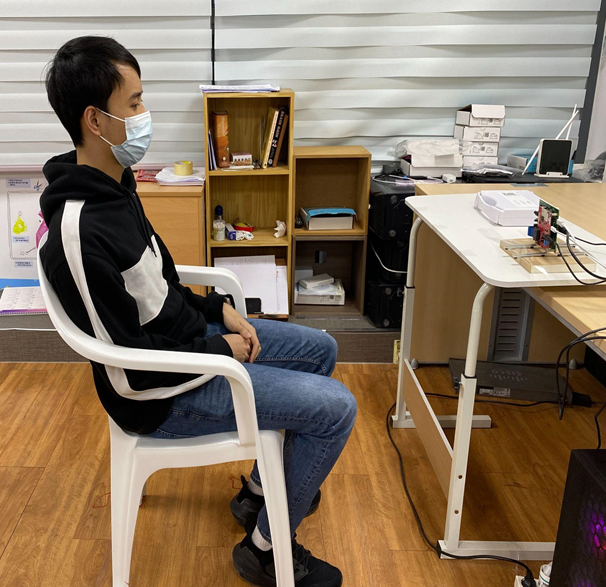} & 
    \includegraphics[width=0.46\linewidth, height=0.36\columnwidth]{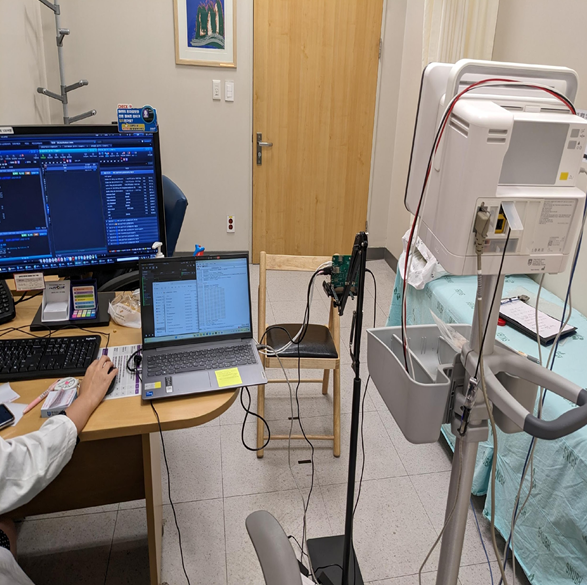} \\ 
    (a) Setup at KAIST & (b) Setup at SNU Hospital \\
  \end{tabular}
  \caption{Experimental setups for data collection. (a) Setup at KAIST. (b) Setup at SNU Hospital.}
  \label{fig:exp}
  \Description{}
\end{figure}

\textbf{Participant:} We collected data from 210 participants (ages 32–68), comprising 108 healthy subjects from the general population and 102 patients diagnosed with arrhythmia. Data from healthy subjects was collected at our institution (KAIST), Korea, while data from arrhythmia patients was collected in collaboration with a renowned cardiologist from Seoul National University Bundang Hospital, Korea. The study was conducted under university Institutional Review Board (IRB) approval, ensuring adherence to ethical guidelines. All participants were briefed on the study objectives, provided written consent before participation, and were compensated for their time. They were also instructed to minimize voluntary movement during recording sessions to ensure consistent data acquisition. Representative experimental setups at both locations are shown in Figure~\ref{fig:exp}.

\textbf{Data Collection Procedure:} The data collection process involved the simultaneous acquisition of radar and ECG recordings from each participant. This was conducted in a typical office environment that included standard furniture and ambient wireless signals such as WiFi, LTE, and Bluetooth, simulating real-life conditions. During data collection, the radar was positioned to capture reflections directly from the participant’s chest, as shown in Figure~\ref{fig:exp}. At the same time, ECG data was collected using the shimmer ECG, with electrodes placed on four locations: the left arm (LA), right arm (RA), left leg (LL), and right leg (RL). Both healthy participants and arrhythmia patients were seated 25 to 55 cm away from the radar. Participants were instructed to breathe normally and minimize voluntary movements to ensure consistent readings.

A total of 210 trials were conducted, each lasting 60 seconds, resulting in a dataset of approximately 20,000 heartbeats. The recorded heart rates ranged from 67 to 97 beats per minute (bpm) for healthy subjects and from 55 to 115 bpm for arrhythmia patients. To ensure the robust performance of the baseline, baseline+PTL, and mCardiacDx systems in reconstructing HPWs and to diagnose arrhythmia, we randomly partitioned the dataset into training, validation, and test sets, following a 60/24/24 split for healthy subjects and 56/22/24 for arrhythmia patients.

\begin{figure}[h]
    \centering
    \includegraphics[width=\linewidth]{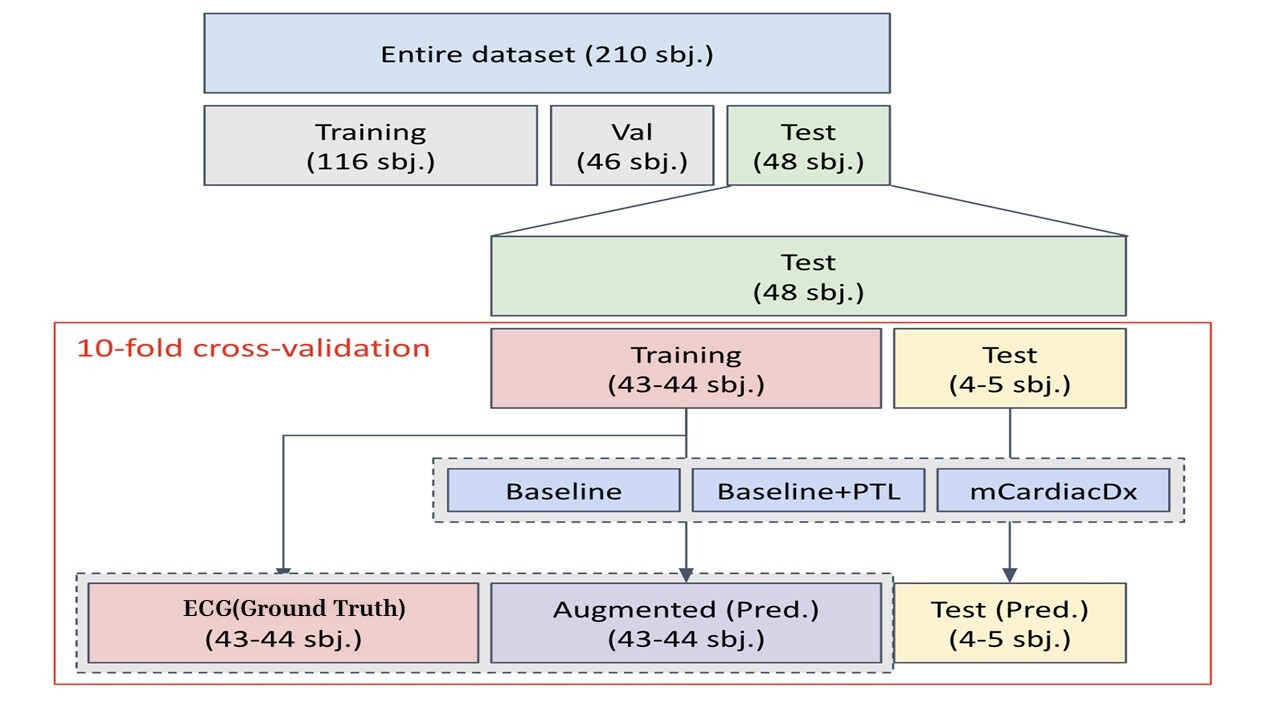}
    \caption{Overview of the dataset split for diagnosis model training and testing.}
    \label{fig:evaluation}
    \Description{}
\end{figure}

\textbf{Network and Model Training: } \label{sec:training} We detail the network and model training below:
\begin{itemize}
    \item  \textit{HPR-Net Training:} The neural networks for the Baseline, Baseline+PTL, and mCardiacDx for HPW reconstruction were trained on the training set for 300 epochs using an Adam optimizer. For the Baseline and Baseline+PTL models, training was performed using a batch size of 16 and an initial learning rate of 0.001. Similarly, the mCardiacDx model was trained using a batch size of 16 and an initial learning rate of 0.0005.

    \item \textit{Diagnosis:} The random forest model was trained on a combined feature set of HRV metrics. This set included metrics derived from ground truth ECG. These ECG-derived metrics were augmented by additional similar HRV metrics derived from the HPWs generated by the three systems (Baseline, Baseline+PTL, and mCardiacDx), as illustrated in Figure~\ref{fig:evaluation}. This approach enables the model to learn directly from both original and system-augmented metrics, which enhance its foundation for robust arrhythmia diagnosis. For testing, model extracts similar metrics derived from the HPWs of the corresponding system.
\end{itemize}

\section{PERFORMANCE RESULTS}
\label{sec:perf}

In this section, we evaluate the performance of Baseline+PTL and mCardiacDx in comparison to the baseline using the evaluation metrics outlined in section~\ref{subsec:evalmetrics}. Our analysis focuses on both healthy subjects and arrhythmia patients, emphasizing three main areas: the quality of HPWs reconstruction, the monitoring of HR and RR intervals, and the diagnosis of arrhythmias.

\subsection{Evaluation Metrics} \label{subsec:evalmetrics}
We employed the following set of quantitative metrics \cite{senin2008dynamic,mdape,sktimeMAPE,novakovic2017evaluation} for each evaluation stage:

\subsubsection{HPW Reconstruction Quality}
\begin{itemize}
    \item \textit{Dynamic Time Warping (DTW):} DTW \cite{senin2008dynamic} measures the similarity between two-time series that may vary in speed or timing. For HPW reconstruction, we use DTW to quantify the similarity between systems HPWs and ECG-generated HPWs. Lower DTW scores indicate higher similarity and better reconstruction.
\end{itemize}

\subsubsection{Heart Rate(HR) and RR Interval Estimation Accuracy}
\begin{itemize}
    \item \textit{Median Absolute Percentage Error (MedAPE):} MedAPE \cite{mdape,sktimeMAPE} quantifies the median absolute percentage difference between estimated values (HR and RR intervals from systems) and ECG. Lower MedAPE values signify higher estimation accuracy.
\end{itemize}

\subsubsection{Arrhythmia Diagnosis Performance}
For evaluating the diagnostic capabilities, we use standard binary classification metrics \cite{novakovic2017evaluation} derived from the confusion matrix, where the arrhythmia is considered the positive class and healthy is the negative class:

\begin{itemize}
    \item \textit{Accuracy:} Proportion of all correctly classified instances, defined as {\small $\text{Accuracy} = (\text{TP} + \text{TN}) / (\text{TP} + \text{FN} + \text{FP} + \text{TN})$}.
    \item \textit{Precision:} Proportion of true positive predictions among all predictions. It measures the model's ability to avoid false positives, defined as {\small $\text{Precision} = \text{TP} / (\text{TP} + \text{FP})$}.
    \item \textit{Recall (Sensitivity):} Proportion of true positive predictions among all actual positive instances. It measures the model's ability to identify all relevant cases, avoiding false negatives, defined as {\small $\text{Recall} = \text{TP} / (\text{TP} + \text{FN})$}.
    \item \textit{F1-score:} The harmonic mean of precision and recall, balancing both. It is a key metric in arrhythmia diagnosis, calculated as {\small $\text{F1-score} = 2 \times (\text{Precision} \times \text{Recall}) / (\text{Precision} + \text{Recall})$}.
    \item \textit{Receiver Operating Characteristic (ROC):} Measures the ability of a classifier to distinguish classes across various threshold settings. A ROC value of 1.0 represents a perfect classifier, while 0.5 indicates a random chance.
\end{itemize}

\subsection{Reconstructing the Heart Pulse Waveform } \label{subsec:hpu}
\begin{figure}[h]
  \centering
  \begin{tabular}{cc}
    \includegraphics[width=0.46\linewidth, height=0.40\columnwidth]{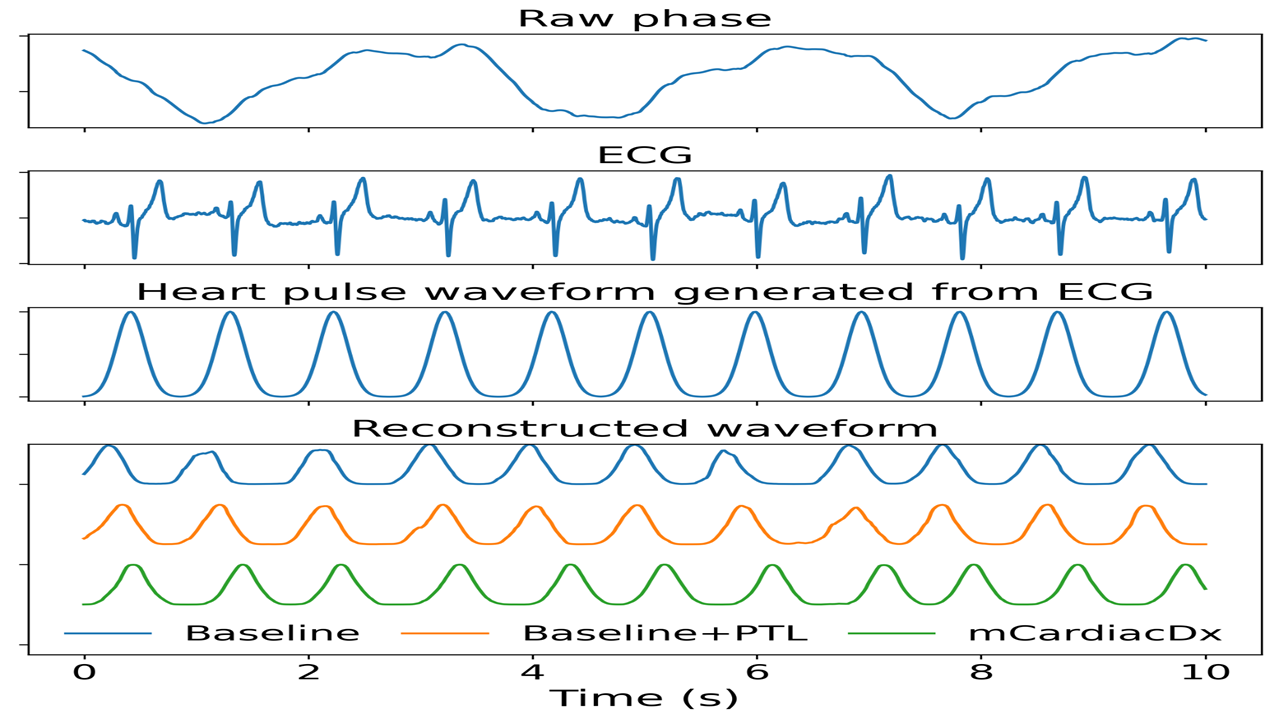} & 
    \includegraphics[width=0.46\linewidth, height=0.40\columnwidth]{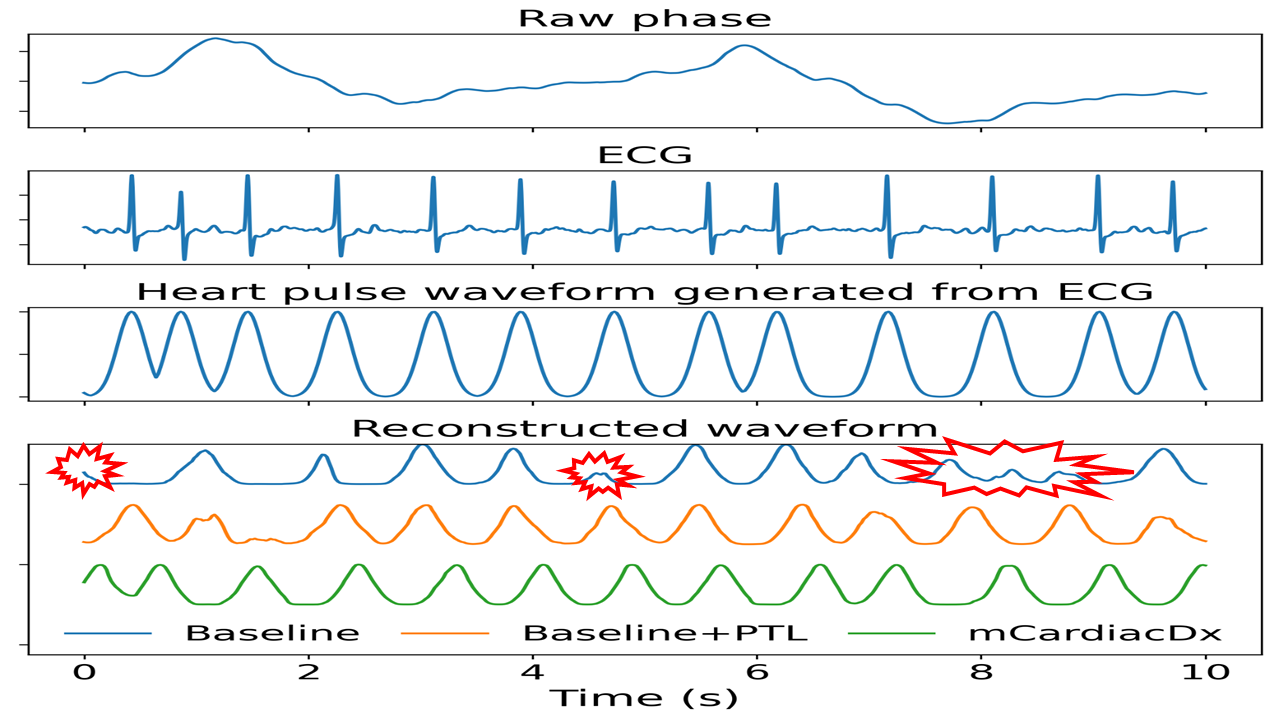}\\ 
    (a) Healthy subjects & (b) Arrhythmia patients \\
  \end{tabular}
  \caption{Reconstructed HPWs of mCardiacDx and Baseline+PTL compared to the Baseline.}
  \label{fig:waveform_both}
  \Description{}
\end{figure}

We evaluate the performance of Baseline+PTL and mCardiacDx in reconstructing HPWs for healthy subjects and arrhythmia patients. Our primary focus is on evaluating how well these systems transform reflection samples into HPWs that closely align with those generated from the ECG compared to the Baseline. Figure~\ref{fig:waveform_both} presents the raw radar phase, ground truth ECG (ECG), HPW generated from the ECG, and HPWs reconstructed by Baseline, Baseline + PTL, and mCardiacDx. Our results in Figure~\ref{fig:waveform_both}(a) show that both Baseline+PTL and mCardiacDx successfully reconstruct HPWs for healthy subjects, which align well with the HPWs generated from ECG compared to the Baseline. However, our primary focus is on arrhythmia patients, where the Baseline initially failed, as highlighted with red area in Figure~\ref{fig:waveform_both}(b).

\begin{figure}[h]
  \centering
  \includegraphics[width=0.58\linewidth, height=0.40\columnwidth]{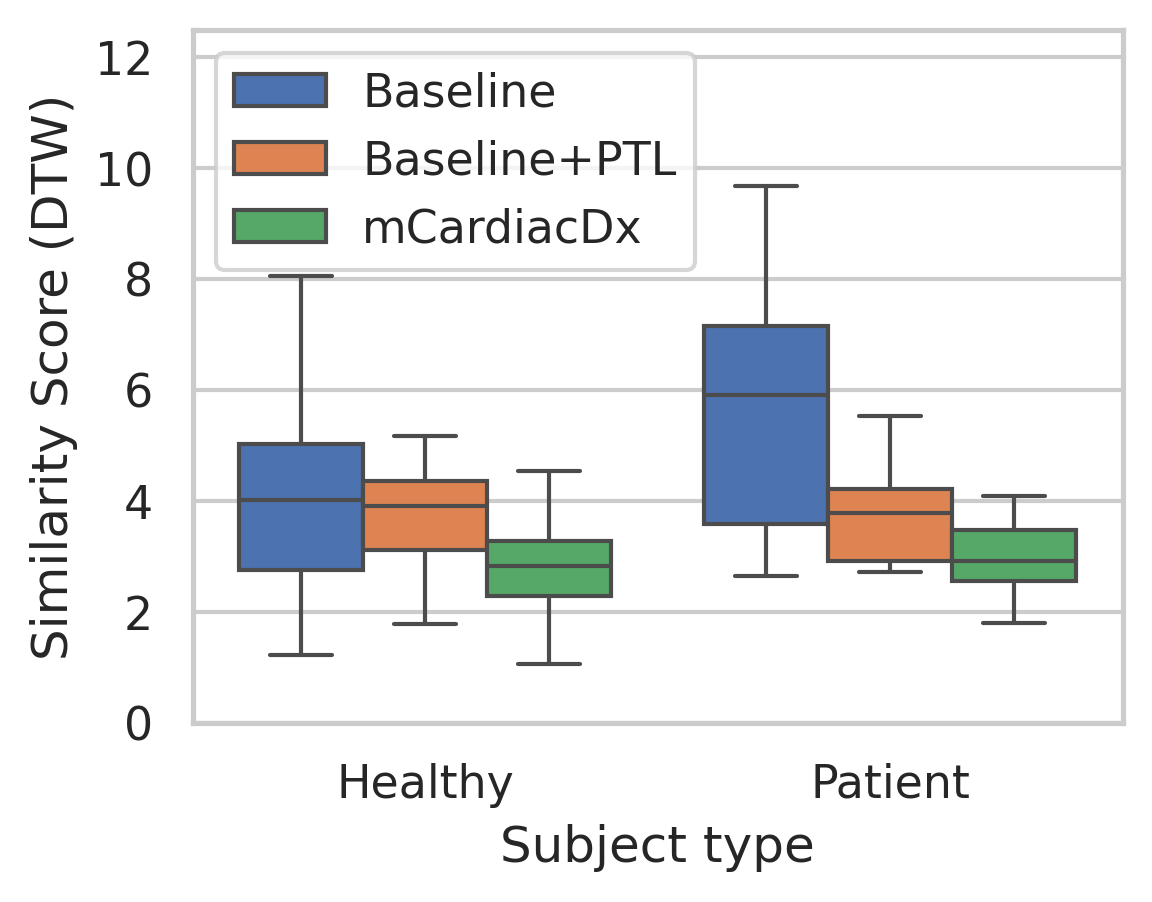} 
  \caption{Dynamic Time Warping (DTW) scores of mCardiacDx and Baseline+PTL compared to the Baseline}
  \label{fig:dtw}
  \Description{}
\end{figure}
To further validate these findings, we perform a similarity analysis between the reconstructed HPWs and those generated from the ECG using dynamic time warping (DTW). DTW is particularly suitable for measuring the similarity or alignment between two-time series that may vary in speed or timing, making it ideal for comparing HPWs in our analysis. The similarity results in Figure~\ref{fig:dtw} show that Baseline + PTL and mCardiacDx achieve strong alignment with the HPW of the ECG, with lower DTW scores of 3.90 and 2.82 for healthy subjects and 3.78 and 2.92 for arrhythmia patients, respectively, compared to baseline scores of 4.02 and 5.92. Given that lower DTW values indicate greater similarity, these results highlight the effectiveness of Baseline+PTL and mCardiacDx in reconstructing HPW from reflection samples closely matching with HPWs generated from ECG, particularly in challenging cases involving arrhythmia where the Baseline tends to falter.

\subsection{Monitoring Arrhythmia} \label{subsec:monitoring}
\begin{figure}[h]
  \centering
  \begin{tabular}{cc}
    \includegraphics[width=0.46\linewidth, height=0.40\linewidth]{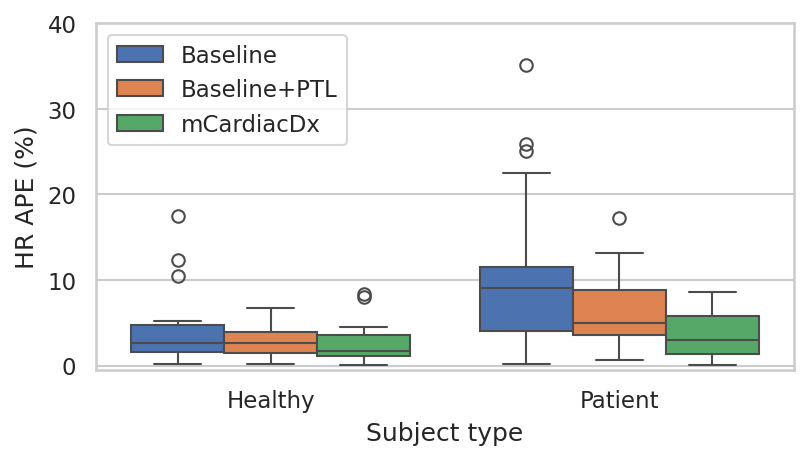} & 
    \includegraphics[width=0.46\linewidth, height=0.40\linewidth]{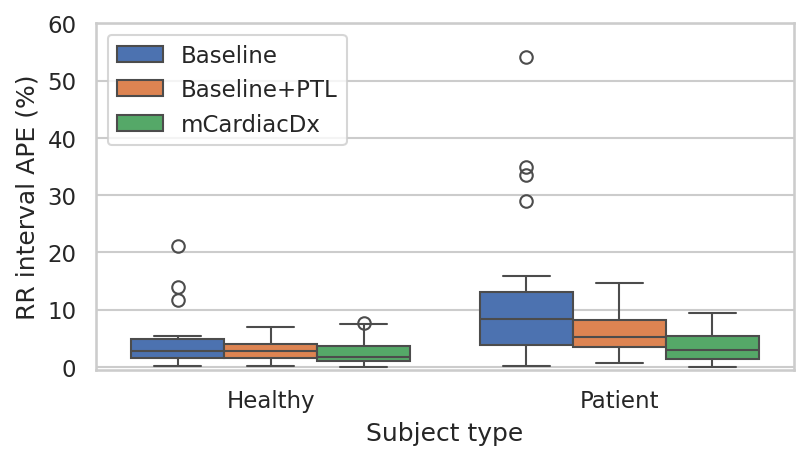} \\ 
    (a) MedAPE of HR & (b) MedAPE of RR Intervals \\
  \end{tabular}
  \caption{Monitoring Performance of mCardiacDx and Baseline+PTL compared to the Baseline.}
  \label{fig:overall_pred_hr_rr}
  \Description{}
\end{figure}

We evaluate the monitoring performance of Baseline+PTL and mCardiacDx in estimating HR and RR intervals based on the reconstructed HPWs. We measure their performance using the MedAPE, calculated relative to ground truth ECG, and compare it against the Baseline method. Figure~\ref{fig:overall_pred_hr_rr} illustrates that both Baseline+PTL and mCardiacDx consistently achieve lower MedAPE error rates for HR and RR intervals estimation across both healthy and arrhythmia subject groups. For the healthy group, Baseline+PTL achieves error rates of 2. 63\% and 2. 70\% for HR and RR intervals, showing a reduction compared to Baseline errors of 2.66\% and 2.73\%, respectively. mCardiacDx further reduces these errors to 1.68\% and 1.71\% for HR and RR intervals. For the arrhythmia group, Baseline+PTL yielded error rates of 4.95\% and 5.21\% for HR and RR intervals, which are significantly lower than the Baseline errors of 9.10\% and 8.42\%. mCardiacDx further reduced these errors to 2.94\% and 2.95\% for HR and RR intervals. These results strongly indicate that both Baseline + PTL and mCardiacDx are robust in monitoring arrhythmias.

\subsection{Diagnosing Arrhythmia} 
\label{subsubsec:diagnosis}
We further evaluate the diagnostic capabilities of Baseline+PTL and mCardiacDx, comparing their performance to the Baseline approach in diagnosing arrhythmia. The findings are presented in the subsequent subsections.

\begin{figure}[h]
  \centering
  \begin{tabular}{cc}
    \includegraphics[width=0.46\linewidth, height=0.40\columnwidth]{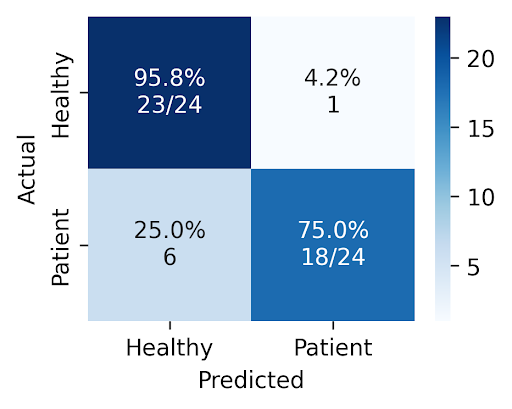} & 
    \includegraphics[width=0.46\linewidth, height=0.40\columnwidth]{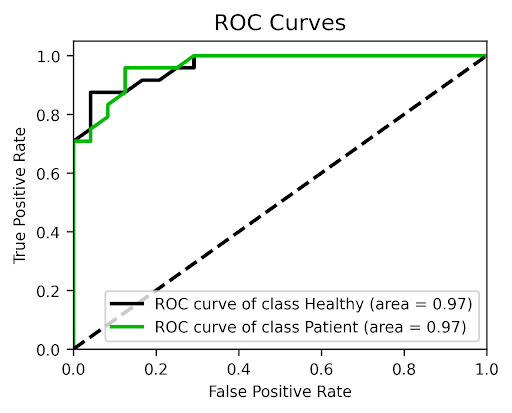} \\ 
    (a) Confusion Matrix & (b) ROC Curves \\
  \end{tabular}
  \caption{Diagnosis Performance of the Baseline.}
  \label{fig:baseline_confusion_roc}
  \Description{}
\end{figure}

\textbf{Baseline:} 
It demonstrates considerable diagnostic performance for healthy subjects, achieving a precision of 0.94 with 1 false positive. However, its recall was limited to 0.7500, resulting in 6 false negatives among the 24 arrhythmia patients, indicating significant challenges in diagnosing arrhythmia cases. While it correctly identifies 23 out of 24 healthy cases and an overall accuracy of 0.8500, the F1-score of 0.83 reflects a moderate balance between precision and recall. The ROC of 0.97 (actual 0.9661) indicates good discriminative ability, though the model's limitations in diagnosing arrhythmia patients prompt further improvements through PTL. The results are illustrated in the confusion matrix and ROC curve plots in Figure~\ref{fig:baseline_confusion_roc}(a) and (b).

\begin{figure}[h]
  \centering
  \begin{tabular}{cc}
    \includegraphics[width=0.46\linewidth, height=0.40\columnwidth]{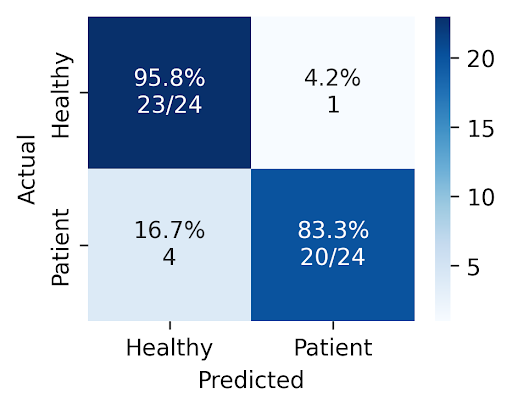} & 
    \includegraphics[width=0.46\linewidth, height=0.40\columnwidth]{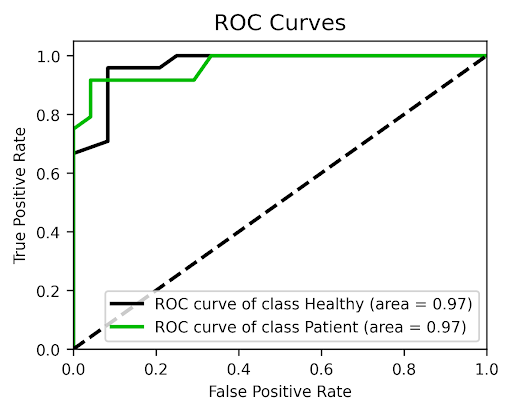} \\ 
    (a) Confusion Matrix & (b) ROC Curve \\
  \end{tabular}
  \caption{Diagnosis Performance of the Baseline+PTL.}
  \label{fig:ptl_confusion_roc}
  \Description{}
\end{figure}

\textbf{Baseline+PTL:}
The Baseline+PTL shows improvements over the Baseline. Its precision and recall increase to 0.95 and 0.83, respectively, with 1 false positive consistent with the Baseline. It diagnoses 20 true positives and reduces false negatives to 4, demonstrating improved sensitivity. While correctly identifying 23 out of 24 true negatives, it achieves an improved F1-score of 0.88, indicating better balance between precision and recall. The overall accuracy increases to 0.89 and the ROC reaches 0.97 (actual 0.9679), demonstrating strong discriminative capacity. Despite these advancements, the remaining false negatives indicate room for improvement, which is addressed by our system mCardiacDx. Nonetheless, Baseline+PTL proves to be a more effective solution for diagnosing arrhythmia compared to the Baseline. The results are shown in the confusion matrix and ROC curve plots in Figure~\ref{fig:ptl_confusion_roc}(a) and (b).

\begin{figure}[h]
  \centering
  \begin{tabular}{cc}
    \includegraphics[width=0.46\linewidth, height=0.40\columnwidth]{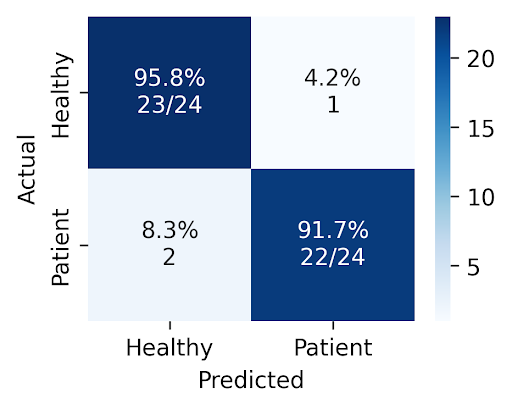} & 
    \includegraphics[width=0.46\linewidth, height=0.40\columnwidth]{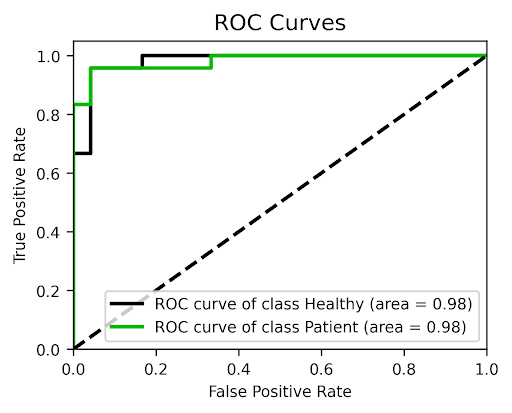} \\ 
    (a) Confusion Matrix & (b) ROC Curve \\
  \end{tabular}
  \caption{Diagnosis Performance of the mCardiacDx.}
  \label{fig:mCardiacDx_confusion_roc}
  \Description{}
\end{figure}

\textbf{mCardiacDx:} 
The mCardiacDx further advances the diagnostic capabilities, achieving a precision of 0.95 with 1 false positive consistent with the Baseline. Its recall improves significantly to 0.91, diagnosing 22 true positives with only 2 false negatives, thus outperforming both the Baseline and Baseline+PTL in sensitivity. While correctly identifying 23 out of 24 true negatives, mCardiacDx demonstrates superior performance with an F1-score of 0.93 and overall accuracy of 0.93. The ROC improves to 0.98, affirming its outstanding discriminative capability. This performance positions mCardiacDx as a leading solution for contactless arrhythmia monitoring. The results are shown in the confusion matrix and ROC curve plots in Figure~\ref{fig:mCardiacDx_confusion_roc}(a) and (b).

\section{Limitations and Future Work}\label{sec:discussion}
While the proposed \textbf{mCardiacDx} system and PTL technique show promise in monitoring and diagnosing arrhythmia, the current implementation is limited to patients in an idle state. This constraint presents challenges for accurately monitoring and diagnosing patients during motion. Moreover, the system currently focuses on reconstructing HPWs, which, although effective, provide limited insight into detailed cardiac dynamics. 

Future work will aim to extend the system and technique to support monitoring and diagnosis under motion. This includes developing motion-robust signal processing techniques to manage distortion and variability introduced by movement. Furthermore, we plan to extend capability of \textit{mCardiacDx} to reconstruct detailed cardiac cycles, enabling analysis comparable to conventional ECG in terms of atrial and ventricular activity. Another important direction involves investigating the relationship between temporally misaligned reflections and arrhythmia severity.

\section{CONCLUSION} \label{sec:conclusion}
In this paper, we present mCardiacDx, a radar-driven contactless system designed for arrhythmia monitoring and diagnosis. The system leverages a PTL technique to analyze reflected signals and reconstruct HPWs for estimating HR, RR intervals, and HRV, which are essential for this purpose. Evaluation results demonstrate that mCardiacDx, powered by PTL, outperform state-of-the-art solution in monitoring and diagnosing arrhythmia, while also demonstrate improved performance in healthy subjects.
\bibliographystyle{ACM-Reference-Format}
\bibliography{references}
\end{document}